\documentclass[amsmath,amsfonts,amssymb]{revtex4}
\usepackage{graphicx}
\usepackage{epsfig}

\newcommand{\non}{\nonumber}
\renewcommand{\a}{\alpha}

\newcommand{\s}{\sigma}
\newcommand{\ts}{\tilde{\sigma}}
\newcommand{\arctanh}{{\rm arctanh}}
\newcommand{\comment}[1]{}
\newcommand{\Tr}{{\rm Tr}}
\newcommand{\eqd}{\stackrel{\text{d}}{=}}

\begin{document}
\title{Replica Cluster Variational Method}

\author{Tommaso Rizzo$^{1,3}$, A. Lage-Castellanos$^{2}$,
  R. Mulet$^{2}$ and Federico Ricci-Tersenghi$^{3}$}

\affiliation{
  $^{1}$ ``E. Fermi'' Center, Via Panisperna 89 A, Compendio Viminale,
  00184, Roma, Italy \\
  $^{2}$ `Henri-Poincar\'e-Group'' of Complex Systems and Department of
  Theoretical Physics, Physics Faculty, University of Havana, La
  Habana, CP 10400, Cuba\\
  $^{3}$ Dipartimento di Fisica, INFN Sezione di Roma1 and CNR-INFM,
  Sapienza Universit\`a di Roma, P.le Aldo Moro 2, 00185 Roma, Italy
}

\begin{abstract}
  We present a general formalism to make the Replica-Symmetric and
  Replica-Symmetry-Breaking ansatz in the context of Kikuchi's Cluster
  Variational Method (CVM).  Using replicas and the message-passing
  formulation of CVM we obtain a variational expression of the
  replicated free energy of a system with quenched disorder, both
  averaged and on a single sample, and make the hierarchical ansatz
  using functionals of functions of fields to represent the messages.
  We obtain a set of integral equations for the message
  functionals. The main difference with the Bethe case is that the
  functionals appear in the equations in implicit form and are not
  positive definite, thus standard iterative population dynamic
  algorithms cannot be used to determine them.  In the simplest cases
  the solution could be obtained iteratively using Fourier transforms.

  We begin to study the method considering the plaquette approximation
  to the averaged free energy of the Edwards-Anderson model in the
  paramagnetic Replica-Symmetric phase.  In two dimensions we find
  that the spurious spin-glass phase transition of the Bethe
  approximation disappears and the paramagnetic phase is stable down
  to zero temperature on the square lattice for different random
  interactions.  The quantitative estimates of the free energy and of
  various other quantities improve those of the Bethe approximation.
  The plaquette approximation fails to predict a second-order
  spin-glass phase transition on the cubic 3D lattice but yields good
  results in dimension four and higher.  We provide the physical
  interpretation of the beliefs in the replica-symmetric phase as
  disorder distributions of the local Hamiltonian.  The messages
  instead do not admit such an interpretation and indeed they cannot
  be represented as populations in the spin-glass phase at variance
  with the Bethe approximation.

  The approach can be used in principle to study the phase diagram of
  a wide range of disordered systems and it is also possible that it
  can be used to get quantitative predictions on single samples. These
  further developments present however great technical challenges.
\end{abstract} 

\maketitle

\section{Introduction}

In the last decade two important results have appeared in the context
of Spin-Glass Theory and disordered systems.  In \cite{MP1} the
formulation of the Replica-Symmetry-Breaking (RSB) ansatz in terms of
populations of fields for the Bethe lattice was presented. This has
led both to the possibility of obtaining new analytical predictions in
the low temperature phase of the model and to the introduction of the
Survey-Propagation (SP) algorithm that has been applied successfully
to random $K$-SAT instances \cite{MP2,MZ,MZP}.  On the other hand it
was recognized that the well-known Belief-Propagation algorithm
corresponds to the Bethe approximation \cite{KS}, and the Generalized
Belief Propagation (GBP) algorithm was subsequently introduced in
Ref.~\cite{YFW,YFW2} as a message-passing algorithm to minimize the
Kikuchi free energy, a more complex approximation than the Bethe one
that goes also under the name of Cluster Variational Method (CVM)
\cite{CVM}\footnote{The original name was ``Cluster {\it Variation}
  Method'', but we believe ``Cluster Variational Method'' to be a
  better wording.}.

Since then the idea of merging these two approaches has been around
but has not been developed so far, probably due the fact that the
standard understanding of the hierarchical ansatz at the Bethe level
comes from the Cavity approximation, while a Cavity-like understanding
of more complex Kikuchi approximations is lacking.  In this paper we
will show that a cavity-like understanding of CVM, although desirable,
is not necessary to implement Replica-Symmetry (RS) and RSB in the
CVM, everything can be worked out using the replica method.

The main idea is to apply the cluster variational method to the
replicated free energy and then to use the RS ansatz or the more
general Parisi's hierarchical ansatz in order to send the number of
replicas $n$ to zero.  We will use the message passing GBP formulation
of CVM representing the messages as populations of populations of
local fields and obtain a set of equations that represents the
generalization of the Survey-Propagation equations.  In principle the
method can be implemented to compute thermodynamic quantities both
averaged over different disorder realizations and on a single sample.
The main difference with the Bethe case is that the populations appear
in the equations in implicit form and therefore standard iterative
population dynamic algorithms cannot be used to determine them.  In
the simplest cases the solution could be obtained iteratively using
Fourier transforms. In appendix \ref{inversion} we will show that, in
principle, the problem may be solved at any level of RSB using
appropriate integral transforms.  Nonetheless, it turns out to be very
hard the actual implementation of this scheme beyond the RS averaged
case or 1RSB on a single sample.  Furthermore the application on
specific models tells us that the messages should not be represented
by populations but rather by functions that are not positive definite
and this represents another technical difficulty, together with the
fact that the equations are written in terms of integrals in many
dimensions.  In general we expect the technical difficulty of the
various approximations to grow very rapidly as we increase the size of
the maximal CVM regions and the number of RSB steps.

We will present the approach in full generality, {\it i.e.}\ for any
CVM approximation and for any number $K$ of RSB steps either with or
without disorder averaging.  The general presentation is somewhat
formal and will be postponed to appendices \ref{RSBparametrization}
and \ref{GSP} instead we will initially discuss the application of the
method to the classic plaquette approximation of the averaged free
energy of the Edwards-Anderson model in the paramagnetic
Replica-Symmetric phase.  In section \ref{plaquette} we write down the
RS message-passing equations for the model and we discuss some of
their features, in particular the misleading analogy between the GBP
equations on a single sample and the RS equations of the averaged
system.  In section \ref{paramagnetic} we compute the free energy and
find that the spurious spin-glass phase transition of the Bethe
approximation disappears and the paramagnetic phase is
thermodynamically stable down to zero temperature for different kinds
of random interactions.  We consider also the 2D triangular and
hexagonal lattices. In both cases the paramagnetic solution yields
positive entropy down to zero temperature, however the triangular
lattice with bimodal interactions has a spurious spin-glass transition
at $T=1.0$, much smaller than the Bethe approximation result $T=2.07$.
We also show that the quantitative estimates of the free energy and of
various other quantities improve those of the Bethe approximation both
qualitatively and quantitatively.  In section \ref{sgtran} we obtain
the location of a possible second-order spin-glass phase transition
studying the Jacobian of the variational equations around the
paramagnetic solution.  This approach predicts no second-order phase
transition on the 2D square lattice, in perfect agreement with the
most reliable numerical studies \cite{HY01,CV6}. Unfortunately it
appears that such a good performance in 2D spoils the 3D result: the
plaquette approximation fails to predict a second-order spin-glass
phase transition on the cubic 3D lattice, which is well seen in
numerical studies \cite{EA3D}.  The same plaquette approximation
provides very good results in dimension four and higher.  Although we
do not solve the equations in the spin-glass phase the analysis of the
Jacobian provides important information on the behaviour of the
messages below the critical temperature.  Most importantly we find
that the messages should not be represented by populations of fields
but rather by functions that are not positive definite.  This
unexpected feature of the actual solutions pushed us to investigate
the physical meanings of the various objects involved in the
computation.  In section \ref{physint} we provide the physical
interpretation of the beliefs in the replica-symmetric phase as the
distributions of the local Hamiltonians with respect to different
realizations of the disorder.  The messages instead do not admit such
an interpretation and therefore need not to be represented as
populations in the spin-glass phase at variance with the Bethe
approximation. We will also discuss the relationship between our
approach and the earlier results of the Tohoku group \cite{T1,T2,T3}.

The present approach can be used in principle to study the phase
diagram of a wide range of disordered systems and it is also possible
that it can be used to get quantitative predictions on single
samples. These further developments present however great technical
challenges and in the last section of the paper we discuss some of
them.

\section{The Replica Approach}
\label{TheReplicaApproach}

The average free energy per spin of a spin-glass system of size $N$ is
computed through the replica method as \cite{MPV}:
\begin{equation}
f=\lim_{n \rightarrow 0} -\frac{1}{\beta n \, N}\ln \langle Z^n_J \rangle
\end{equation}
Where $J$ label different realizations of the disorder and the angular
brackets mean average over them.  In recent years it has been realized
that the Replica-Symmetry-Breaking solution can be usefully applied
(through the cavity method) to a given disorder realization, in the
replica framework this corresponds to write the free energy as:
\begin{equation}
f_J=\lim_{n \rightarrow 0} -\frac{1}{\beta n \, N}\ln Z^n_J
\end{equation}
The above expression is apparently trivial because the replicas are
uncorrelated if we do not average over the disorder, however in the
spin-glass phase the true thermodynamic state is the one in which the
distinct replicas are actually correlated because of spontaneous RSB.
In the replicated Edwards-Anderson model we can define the following
functional:
\begin{equation}
  \Phi(n) = -{1\over n \beta \, N}\ln \Tr \langle \exp(\sum_{(ij)} \beta
  J_{ij} \sum_{a=1}^n s_i^a s_j^a)\rangle = -{1\over n\beta \, N}\ln \Tr
  \exp\left(\sum_{(ij)}\ln \langle \exp \beta J \sum_a s_i^a s_j^a
    \rangle \right)\;,
\end{equation}
such that the free energy is obtained as the $n \rightarrow 0$ limit
of the above expression.  For a single sample the analogous function
$\Phi_J(n)$ is obtained removing the averages over the couplings
$J_{ij}$.  Although in the present paper we will be interested in the
$n \to 0$ limit, it is worth noticing that the replica cluster
variational method can provide an approximation to the entire function
$\Phi(n)$, which is related to free energy fluctuations from sample to
sample \cite{CPSV, PR08}.  Moreover, at the RS level of approximation,
the value $\max_n \Phi(n)$ may provide a better approximation to the
typical free-energy than $\Phi(0)$ (at the cost of introducing
reweighting terms in the RS integral equations).

In the following we will consider regions of spins $r$ and we will use
the definition $\psi_r(\s_r)=\prod_{i,j \in r }\langle \exp \beta J
\sum_a s_i^a s_j^a \rangle$ (or $\psi_r(\s_r)=\prod_{i,j \in r } \exp
\beta J \sum_a s_i^a s_j^a $ on a given sample) to make contact with
the notations of Ref.~\cite{YFW}. The difference between the two cases
is that in the averaged case $\psi_r(\s_r)$ is homogeneous over space
while it fluctuates on a single sample. We will refer to the region of
two interacting spins as $ij$.

In both cases the functional $\Phi(n)$ can be regarded as the
equilibrium free energy of a replicated model.  The free energy can be
expressed through a variational principle. The resulting expression
involves an energetic and an entropic term. The problem is that the
latter is usually very difficult to be treated. A standard way to
treat it is Kikuchi's cluster variational method. In its modern
formulation this method consists in writing the entropic term as a sum
of entropic cumulants and then to truncate this expansion at some
order, see \cite{PEL} for a recent detailed presentation.

Basically the starting point of the approximation is to chose a set of
regions of the graph over which the model is defined. These are called
the maximal clusters and the entropic expansion is truncated assuming
that the cumulants of larger regions vanish. In recent years it has
been realized that the variational equations can be written in a
message passing way \cite{YFW} and we will use this formulation in
order to extend the CVM to replicated models, either averaged or not
over the disorder.

\section{Cluster Variational Method and Message-Passing}
\label{CVMmp}

In the following we will briefly present the message-passing approach
to cluster variational method of Ref.~\cite{YFW}. We will use the same
notation of Ref.~\cite{YFW} and we refer to it for a more detailed
presentation.

We will call $R$ a set of connected clusters (regions) of nodes
(spins), plus their intersections, plus the intersections of the
intersections and so on. Then $x_r$ is the state (configuration) of
nodes in $r$ and $b_r(x_r)$ (the belief) is an estimate of the
probability of configuration $x_r$ according to the Gibbs measure.
Following \cite{YFW} we will often use the notation $b_r$ omitting the
explicit dependence of the beliefs $b_r(x_r)$ on $x_r$. Then the
energy of region $r$ is $E_r=-\ln \prod_{ij}\psi_{ij}(x_i,x_j)-\ln
\prod_i \psi_i(x_i)$ where the products run over all links and nodes
(in presence of a field) contained in region $r$. With this
definitions, the Kikuchi free energy reads:
\begin{equation}
F_{K}=\sum_{r \in R}c_r \left( \sum_{x_r}b_r E_r+\sum_{x_r}b_r \ln b_r \right)
\label{Fbeliefs}
\end{equation}
where the so-called Moebius coefficient $c_s$ is the over-counting
number of region $s$ defined by $c_s=1-\sum_{r \in {\cal
    A}(s)}c_r$. The set ${\cal A}(s)$ is made of all ancestors of $s$,
{\it i.e.}\ it is the set of all regions that contain $s$. The
condition $c_s=1$ holds for the largest regions.

The cluster variational method amounts to extremize the free energy
with respect to the beliefs, under the constraint that the beliefs are
normalized and compatible one with each other in the sense that the
belief of a region can be obtained marginalizing the belief of any of
its ancestors.  It is worth noticing that the Kikuchi free energy does
not provide in general an upper bound on the true free energy of the
model.

The main result of Ref.~\cite{YFW} was to show that the variational
equations for the beliefs can be written in a message-passing
fashion. In order to do this we define for any given region $r$: i)
the set of its ancestors ${\cal A}(r)$, that is the set of regions
that contain region $r$; ii) the set of its parents ${\cal P}(r)$,
that is the subset of its ancestors that have no descendant that is
also an ancestor of $r$; iii) the set of its descendants ${\cal
  D}(r)$, that is the set of regions contained in region $r$; iv) the
set of its children ${\cal C}(r)$, that is the subset of its
descendants that are not contained in a region that is also a
descendant of $r$.  One introduces message $m_{rs}$ from a region $r$
to any of its children $s$.  The messages can be thought of as going
from the variable nodes (spins) in $r\setminus s$ to the variable
nodes in $s$.  They depend on the configuration of $x_s$ but for
brevity this dependence is omitted.  We also need the following
definitions \footnote{We adopt the original notation of
  Ref.~\cite{YFW}, which was changed in the more recent
  Ref.~\cite{YFW2}. The ensembles $M(r)\setminus M(s)$ and $M(r,s)$
  corresponds respectively to $N(r,s)$ and $M(r,s)$ defined in
  \cite{YFW2}. Note however that for us these are ensembles of couples
  of regions labels instead of ensembles of the corresponding messages
  as in \cite{YFW}.}:
\begin{itemize}
\item $M(r)$ is defined as the ensemble of connected (parent-child)
  pairs of regions $(r',s')$ such that $r'\setminus s'$ is outside $r$
  while $s'$ coincides either with $r$ or with one of its descendants.
\item $M(r)\setminus M(s)$ is the ensemble of connected pairs of
  regions that are in $M(r)$ but not in $M(s)$.
\item $M(r,s)$ is the ensemble of connected pairs of regions such that
  the parent is a {\it descendant} of $r$ and the child is either
  region $s$ or a descendant of $s$.
\end{itemize}
Although all these definitions of sets of regions may look abstract
and hard to follow, in the next Section we will provide immediately an
example on the 2D square lattice which should make these definitions
clearer.

With the above definitions it can be shown \cite{YFW,YFW2} that the
beliefs can be expressed as:
\begin{equation}
b_r=\alpha_r \psi (x_r) \prod_{r's'\in M(r)} m_{r' s'}
\label{belief}
\end{equation}
where $\alpha_{r}$ is a normalization constant because they are
probability distributions.  The messages obey the following equations:
\begin{equation}
  m_{rs} = \alpha_{rs} \sum_{x_{r \setminus  s}} \psi_{r\setminus
    s}(x_{r}) \prod_{{r'' s''} \in M(r)\setminus M(s)}m_{r''
    s''}/\prod_{{r' s'} \in M(r,s)} m_{r' s'}
\label{yedmessage}
\end{equation} 
where $\alpha_{rs}$ is some constant (see below) and $\psi_{r\setminus
  s}(x_r)$ is the set of interactions between the nodes of region $r$
without considering those that are just in region $s$, {\it i.e.}\
$\psi_{r\setminus s}(x_r) \equiv \psi_{r}(x_r)/\psi_s(x_s)$.

The constants in eq. (\ref{yedmessage}) can be fixed to any positive
value, indeed the messages need not to be normalized. In \cite{YFW2}
the constants $\alpha_{rs}$ are fixed to 1, while here, for reason of
convenience, \emph{we work with messages normalized to unity}, and the
$\alpha_{rs}$ have to be intended as normalization constants.  Two
sets of messages obtained solving the equation with two different sets
of values of the constants $\alpha_{rs}$ are simply related by
positive multiplicative factors.

In general the Kikuchi free energy has to be extremized with respect
to the beliefs $b_r$ under the constraint that they are compatible, in
the sense that the belief of one region marginalized to one of its
subregion $s$ has to be equal to $b_s$. This is done introducing
appropriate Lagrange multipliers. The results quoted above have been
obtained showing that there exists an equivalent set of constraints
between each parent-child couple $(r,s)$ such that imposing these
constraints through a set of Lagrange multiplier $\mu_{rs}$,
extremizing with respect to the beliefs and identifying $m_{rs}=\exp
\mu_{rs}$, one immediately gets equation (\ref{belief}); equation
(\ref{yedmessage}) for the messages is obtained imposing the standard
marginalization conditions for the beliefs.  This makes clear why the
message $m_{rs}$ as a function of $x_s$ can be normalized to any
positive constant, indeed this corresponds to a constant shift in the
definition of the Lagrange multipliers $\mu_{rs}$.

Once the beliefs are obtained the Kikuchi free energy can be
computed. However for our purposes we derive another expression of the
free energy in terms of the messages.  To do this we note that if one
substitutes the expression for the beliefs in the Kikuchi free energy
plus the Lagrange multipliers terms one obtains the following {\it
  variational expression for the free energy}
\begin{equation}
  F_K= -\sum_{r \in R}c_r \ln \left[ \sum_{x_{r}}\psi_{r} (x_{r})
    \prod_{r's'\in M(r)}m_{r's'}\right]
\label{varf}
\end{equation}
This expression for $F_K$ is stationary with respect to the messages
in the sense that the equations (\ref{yedmessage}) can be also
obtained extremizing it with respect to the messages
\footnote{Actually one obtain a different set of equations that can be
  proved to be equivalent to eqs. (\ref{yedmessage}) along the same
  lines of the proof that the two sets of constraints are equivalent,
  see theorem 5 in \cite{YFW2}}.  The fact that we can choose any
normalization for the messages can be also derived noticing that the
variational free energy is invariant under a rescaling of the messages
$m_{pr} \rightarrow a_{pr}m_{pr}$. Indeed the resulting free energy
would differ by a term
\begin{equation}
  -\left(c_r+\sum_{A\in {\cal A}(r) \setminus  p \cup {\cal P}(p)}
    c_A\right)\ln a_{pr}=0
\end{equation}
as can be seen subtracting the two equations $c_r=1-\sum_{A \in {\cal
    A}(r)}c_A$ and $c_p=1-\sum_{A \in {\cal A}(p)}c_A$, see eq. 124 in
\cite{YFW2}.
  
For later convenience we also introduce the following messages
normalized to unity:
\begin{equation}
  \tilde{m}_{rs} \propto  \,m_{rs}\prod_{{r' s'} \in M(r,s)} m_{r' s'}
\label{tm}
\end{equation}
Accordingly the tilded messages defined above obey an equation with no
messages at the denominator:
\begin{equation}
  \tilde{m}_{rs}=\tilde{\alpha}_{rs} \sum_{x_{r \setminus
      s}}\psi_{r\setminus s} (x_{r})\prod_{r'' s''\in M(r)\setminus
    M(s)}m_{r'' s''} 
\label{tm1}
\end{equation}
In the case of replicated models $x_r$ defines the state of the spins
in regions $r$ where each spin is replicated $n$ times. For any finite
integer $n$ the above equations in principle can be solved explicitly,
but in order to make the analytical continuation to real small $n$ we
need to rephrase them in an appropriate way. This will be done using
the hierarchical ansatz. The hierarchical ansatz was introduced by
G. Parisi in the context of fully-connected spin-glass models
\cite{MPV} and later extended to spin-glasses defined on random
lattices where the Bethe approximation is correct
\cite{MP1,MP2,MZ,MZP}. It is also called the Replica-Symmetry-Breaking
(RSB) ansatz and can be introduced with different levels of RSB steps
$K$.  The value $K=0$ corresponds to the Replica-Symmetric case that
in spin-glasses is assumed to be correct in the paramagnetic phase
valid at high enough temperature or magnetic field. The RS
parametrization is already non-trivial in the present general CVM
context and presents some substantial differences with the Bethe
approximation.

In the following sections we will consider the message
passing-formulation of the CVM plaquette approximation at the RS level
and study its high-temperature phase quantitatively.  The general RSB
for a generic CVM approximation will be presented in the Appendices at
the end of the paper.

\section{The Replica-Symmetric plaquette approximation}
\label{plaquette}

\subsection{Message passing equations}

In this section and in the following we study the plaquette
approximation for the replicated Edwards-Anderson model on various
regular lattices in dimension $D$.  The plaquette approximation is the
oldest improvement on the Bethe approximation \cite{CVM}.  A detailed
presentation of its message-passing formulation can be found in
\cite{YFW} and \cite{PEL}.  In this approximation there are three
regions: the plaquette of four spins, the couple of spins and the
single spin (the point-like region).  To make connection with the more
abstract definition of regions given in the previous Section, please
note that $\mathcal{A}(\text{point}) = \{\text{plaquette, couple}\}$,
$\mathcal{P}(\text{point}) = \{\text{couple}\}$,
$\mathcal{A}(\text{couple}) = \{\text{plaquette}\}$,
$\mathcal{P}(\text{couple}) = \{\text{plaquette}\}$,
$\mathcal{D}(\text{couple}) = \{\text{point}\}$,
$\mathcal{C}(\text{couple}) = \{\text{point}\}$,
$\mathcal{D}(\text{plaquette}) = \{\text{couple, point}\}$,
$\mathcal{C}(\text{plaquette}) = \{\text{couple}\}$, and void sets are
not reported.  Thus we deal with two types of replicated-spins
messages, from plaquette to couple and from couple to point.  If we do
not average over the disorder the messages depends on the position of
the corresponding regions on the lattice while if we average over the
disorder all the replicated messages are the same and we have to deal
with just two of them.

We will work at the RS level, {\it i.e.}\ with $K=0$ RSB
steps. Physically this corresponds to the case where on each sample
there is a single thermodynamic state and the messages are just
numbers that fluctuate over space.  Correspondingly in the averaged
case the messages are functions that are spatially homogeneous as we
will see below.  On a single sample the equations we will obtain
correspond to GBP, while in the averaged case the equations are new.
As we will see in the following, looking at the equations of the
averaged case as if the message functions represent the spatial
distribution of the GBP messages on a single-sample is completely
wrong; we will come back on this issue several times later in the
paper and finally present the correct interpretation of the various
quantities in section \ref{physint}.

\begin{figure}[htb]
\begin{center}
\epsfig{file=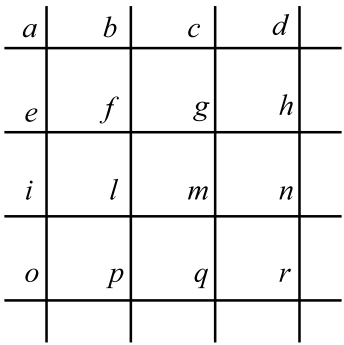,width=6cm}
\caption{A portion of the 2D square lattice. In the plaquette
  approximation we have couple-to-point messages from, say $fg$ to
  $f$, and plaquette-to-couple messages from, say, $fglm$ to $lm$.}
\label{fig2dindex}
\end{center}\end{figure}

Let us start with the equation for a given sample defined on a 2D
square lattice. At the RS level this corresponds to work with a single
replica.  With reference to Fig.~\ref{fig2dindex} and according to
eq. (\ref{yedmessage}) we see that we have two types of messages. The
first type of message is from a couple of spins, say $fg$, to a spin
$f$ and it is a function $\rho_{fg,f}(\sigma_f)$ of the value of the
Ising spin $\sigma_f$.  As a consequence the message can be
parametrized by a single field $u_{fg,f}^f$ according to the following
expression:
\begin{equation}
\rho_{fg,f}(\sigma_f) = \exp[\beta u_{fg,f}^f \sigma_f] / (2
\cosh(\beta u_{fg,f}^f))\;,
\end{equation}
The second type of messages is from a region of four spin, say, $fglm$
to a couple of spins $lm$ and it is a function
$\rho_{fglm,lm}(\sigma_l,\sigma_m)$ of the two Ising spins
$(\sigma_l,\sigma_m)$, as a consequence it can be parametrized by
three fields $(U_{fglm,lm}^{lm},u_{fglm,lm}^l,u_{fglm,lm}^m)$ in the
following way:
\begin{equation}
  \rho_{fglm,lm}(\sigma_l,\sigma_m)=\exp[\beta
  U_{fglm,lm}^{lm}\sigma_l\sigma_m+ \beta u_{fglm,lm}^l\sigma_l+ \beta
  u_{fglm,lm}^m\sigma_m]{\cal
    N}(U_{fglm,lm}^{lm},u_{fglm,lm}^l,u_{fglm,lm}^m) 
\end{equation}
Where ${\cal N}(U_{fglm,lm}^{lm},u_{fglm,lm}^l,u_{fglm,lm}^m)$ is a
normalization constant that enforces the condition
$\sum_{\{\sigma_l,\sigma_m\}}\rho_{fglm,lm}(\sigma_l,\sigma_m)=1$.
Now equations (\ref{yedmessage}) can be rewritten in terms of the
various $u$-fields and yield a closed set of equations for them.  In
order to write down these equations explicitly we introduce the
following functions:
\begin{equation}
  \hat{h}(U,u_1,u_2)=u_1+{1 \over 2 \beta}\ln \frac{\cosh \beta
    (U+u_2)}{\cosh \beta (U-u_2)} 
\end{equation}
and 
\begin{eqnarray}
  \hat{U}(U_{12},U_{23},U_{34},u_1,u_2,u_3,u_4) & = & {1 \over 4
    \beta} \ln {K(1,1)K(-1,-1) \over K(1,-1)K(-1,1)} \\
  \hat{u}_1(U_{12},U_{23},U_{34},u_1,u_2,u_3,u_4) & = & {1 \over 4
    \beta} \ln {K(1,1)K(1,-1) \over K(-1,1)K(-1,-1)} \\
  \hat{u}_2 (U_{12},U_{23},U_{34},u_1,u_2,u_3,u_4) & = & {1 \over 4
    \beta} \ln {K(1,1)K(-1,1) \over K(1,-1)K(-1,-1)} 
\end{eqnarray}
where 
\begin{equation}
  K(\s_{1},\s_{4})=\sum_{\{\s_{2},\s_{3}\}}B(\s_{1},\s_{2},\s_{3},\s_{4}) \ ,
\end{equation}
\begin{equation}
  B(\s_{1},\s_{2},\s_{3},\s_{4})  = \exp \beta [ U_{12}\s_1
  \s_2+U_{23}\s_2 \s_3+U_{34}\s_3 \s_4+u_1 \s_1+u_2 \s_2+u_3 \s_3+u_4
  \s_4] \ .
\end{equation}
The equation for the field $u_{fg,f}^f$ reads:
\begin{equation}
  u_{fg,f}^f=\hat{h}(U_{bcfg,fg}^{fg}+U_{fglm,fg}^{fg}+J_{fg},
  u_{bcfg,fg}^{f}+u_{fglm,fg}^{f},
  u_{bcfg,fg}^{g}+u_{fglm,fg}^{g}+u_{cg,g}^g+u_{mg,g}^g+u_{hg,g}^g) 
\label{arghat}
\end{equation}
where the l.h.s. corresponds to the r.h.s. of equation (\ref{tm}) and
the r.h.s. corresponds to the r.h.s. of eq. (\ref{tm1}).  The equation
for the $2$-field from the plaquette $fglm$ to the couple $lm$ reads:
\begin{eqnarray}
U_{fglm,lm}^{lm} & = & \hat{U}(\#)
\label{GBP1}
\\
u_{fglm,lm}^{l}+u_{fl,l}^l & = & \hat{u}_1(\#)
\label{GBP2}
\\
u_{fglm,lm}^{m}+u_{gm,m}^m & = & \hat{u}_2(\#)
\label{GBP3}
\end{eqnarray}
where the notation $\#$ stands for the fact that we have to substitute
the arguments of the functions $\hat{U}$,$\hat{u}_2$ and $\hat{u}_2$
according to:
\begin{eqnarray}
U_{12} & = & U_{efil,lf}^{lf}+J_{lf}
\label{sommaMessPrimo} 
\\
 U_{23} & = & U_{bcfg,fg}^{fg}+J_{fg}
\\
U_{34} & = & U_{ghmn,gm}^{gm}+J_{gm}
\\
u_1 & = & u_{efil,lf}^{l}
\\
u_2 & = & u_{efil,lf}^{f}+u_{bcfg,fg}^{f}+u_{ef,f}^f+u_{bf,f}^f
\\
u_3 & = & u_{bcfg,fg}^{g}+u_{ghmn,gm}^{g}+u_{cg,g}^g+u_{hg,g}^g
\\
u_4 & = & u_{ghmn,gm}^{m}
\label{sommaMessUltimo} 
\end{eqnarray}
Again the l.h.s. of eqs. (\ref{GBP1},\ref{GBP2},\ref{GBP3}) correspond
to the r.h.s. of eq. (\ref{tm}) while the r.h.s. correspond to the
r.h.s. of eq. (\ref{tm1}).  As we already said up to this point we
have just written the GBP equations of \cite{YFW}.

Now we turn to the replicated CVM and we study it in the average
case. On each site, say $f$, there are $n$ spins $\ts_f \equiv
(\s_f^1, \dots \, , \s_f^n )$ and the replicated spins interact with
their neighbors, say $\ts_g$ with an interaction term of the form
$\psi_{fg}(\ts_f,\ts_g) \equiv \int P(J_{fg})dJ_{fg} \exp[\beta
J_{fg}\sum_{\a=1,n} \s_f^{\a} \s_g^{\a}]$. The general RS and RSB
ansatz of a function $\rho(\ts)$ of $n$ spins was originally presented
in \cite{GDD}, later its parametrization in terms of distributions of
fields was suggested in \cite{Mon1} and later revisited in \cite{MP1},
and we refer to those paper for an explanation of the main ideas
underlying it. We have generalized it to a generic function
$\rho(\ts_1,\dots ,\ts_p)$ where each $\ts_i$ is a set of $n$ Ising
spins, in this section we present the RS case while the general RSB
case is presented in the appendices.

Let us consider the first kind of message, the couple-to-point, say
$\rho_{fg,f}(\ts_f)$. For each integer $n$ we could parametrized it
through $2^n-1$ fields, however such a construction is not suitable to
perform the $n \rightarrow 0$ limit. Therefore following
\cite{GDD,Mon1,MP1} we parametrized it through a message function
$q_{fg,f}(u_{fg,f}^f)$ in the following way:
\begin{equation}
\rho_{fg,f}(\ts_f)=\int dq_{fg,f} \exp[\beta u_{fg,f}^f
\sum_{\alpha=1}^n\sigma_f^{\alpha}] (2\cosh \beta u_{fg,f}^f)^{-n} 
\end{equation}
where we have used the shorthand notation $dq_{fg,f} \equiv
q_{fg,f}(u_{fg,f}^f)du_{fg,f}^f$.  The plaquette-to-couple message in
turn is parametrized through a message function
$Q_{fglm,lm}(U_{fglm,lm}^{lm},u_{fglm,lm}^l,u_{fglm,lm}^m)$ as:
\begin{eqnarray}
\rho_{fglm,lm}(\ts_l,\ts_m) & = & \int dQ_{fglm,lm} {\cal
  N}(U_{fglm,lm}^{lm},u_{fglm,lm}^l,u_{fglm,lm}^m)^n\times 
\nonumber
\\
& \times & \exp[\beta
U_{fglm,lm}^{lm}\sum_{\alpha=1}^n\sigma_l^{\alpha}\sigma_m^{\alpha}+
\beta u_{fglm,lm}^l\sum_{\alpha=1}^n\sigma_l^{\alpha}+ \beta
u_{fglm,lm}^m\sum_{\alpha=1}^n\sigma_m^{\alpha}] 
\end{eqnarray}
where we have used the shorthand notation 
\begin{equation}
dQ_{fglm,lm}\equiv
Q_{fglm,lm}(U_{fglm,lm}^{lm}, u_{fglm,lm}^l, u_{fglm,lm}^m)
dU_{fglm,lm}^{lm}\, du_{fglm,lm}^l\,du_{fglm,lm}^m
\end{equation}
The above parametrization allows to rewrite the message-passing
equations (\ref{yedmessage}) as a set of equations for the message
functions for any replica number $n$. The derivation of these
equations is conceptually straightforward and it is given in the
appendix. Here we just quote the results, in particular in the $n
\rightarrow 0$ we have:
\begin{equation}
q_{fg,f}(u_{fg,f}^f) = \int \delta(u_{fg,f}^f-\hat{h}) dQ_{bcfg,fg}
dQ_{fglm,fg} dq_{cg,g} dq_{gh,g} dq_{gm,g} dP(J_{fg})
\label{plav1}
\end{equation}
where the arguments of the function $\hat{h}$ are as in
eq. (\ref{arghat}), $dP(J_{fg})=P(J_{fg})dJ_{fg}$ and $P(J_{fg})$ is
the disorder distribution of the quenched coupling $J_{fg}$.  The
equation for $Q_{fglm,lm}$ is obtained from eqs. (\ref{tm}) and
(\ref{tm1}). The l.h.s. is given by the r.h.s. of eq. (\ref{tm}) and
reads:
\begin{eqnarray}
{\rm l.h.s.} &  = & \int
\delta(\tilde{U}_{fglm,lm}^{lm}-U_{fglm,lm}^{lm})
\delta(\tilde{u}_{fglm,lm}^l-u_{fglm,lm}^l-u_{fl,l}^l)
\delta(\tilde{u}_{fglm,lm}^m-u_{fglm,lm}^m-u_{gm,m}^m) \times
\nonumber
\\
& \times & dQ_{fglm,lm}dq_{fl,l}dq_{gm,m}\ .
\label{lhs}
\end{eqnarray}
The r.h.s. is given by the r.h.s. of eq. (\ref{tm1}) and reads:
\begin{eqnarray}
{\rm r.h.s.} &  = & \int \delta(\tilde{U}_{fglm,lm}^{lm}-\hat{U}(\#))
\delta(\tilde{u}_{fglm,lm}^l-\hat{u}_1(\#))
\delta(\tilde{u}_{fglm,lm}^m-\hat{u}_2(\#)) \times
\nonumber
\\
& \times & dP(J_{lf})dP(J_{fg})dP(J_{gm}) dQ_{efil,fl} dQ_{bcfg,fg}
dQ_{ghmn,gm}dq_{lf,f}dq_{bf,f}dq_{cg,g}dq_{gh,g}\ .
\label{rhs}
\end{eqnarray}
where the arguments of the functions $\hat{U}$,$\hat{u}_2$ and
$\hat{u}_2$ are as in eqs. (\ref{GBP1},\ref{GBP2},\ref{GBP3}). Now
equating the l.h.s. and r.h.s. written above for each value of the
auxiliary variables
$(\tilde{U}_{fglm,lm}^{lm},\tilde{u}_{fglm,lm}^{l},\tilde{u}_{fglm,lm}^{m})$
we get the equation for $Q_{fglm,lm}$.

The CVM free energy can also be expressed in terms of the various
message functions $Q_{fglm,lm}$ and $q_{fg,f}$. It is presented in
full generality in appendix $\ref{GSP}$ and will not be reported
here. The resulting expression is variational in the sense that the
above equations can be also obtained extremizing it with respect to
its arguments {\it i.e.}\ the various functions $Q_{fglm,lm}$ and
$q_{fg,f}$. The number of replicas $n$ appears as a parameter in the
variational free energy and the analytical continuation to non-integer
values can be performed. As we will see in appendix \ref{GSP}, in
order to derive such an expression it is crucial to start from the
variational expression (\ref{varf}) which depends on the messages and
not on the beliefs as expression (\ref{Fbeliefs}).

\subsection{Discussion}

In the following we will briefly discuss the equations just obtained
for the message functions.  First of all we note that in the average
case the replicated Hamiltonian is spatially homogeneous and therefore
it is natural to assume that all the couple-to-point message functions
are described by a single function $q(u)$ and all the
plaquette-to-couple message functions are described by a single
function $Q(U,u_1,u_2)$.  On the other hand, the above equations gives
us an idea of what it would look like to be working on a single sample
at the 1RSB level. Indeed in this case we should have message
functions fluctuating in space and obeying the above equations
(obviously without the average over the couplings $J$). As we will see
in appendices \ref{RSBparametrization} and \ref{GSP} the above
equations would corresponds to make the 1RSB on a single sample with
$x_1=0$ while in order to treat the general case $x_1>0$ we should add
the appropriate reweighting terms predicted by eqs. (\ref{itesamp1})
and (\ref{itesamp2}).
  
We note two important technical difficulties that one has to face
solving the above equations, also in the averaged case in which one
have to deal with just two integral equations for the functions $q(u)$
and $Q(U,u_1,u_2)$.  First of all they involve integrals in many
dimensions, {\it e.g.}\ eq. (\ref{rhs}) requires in principle the
computation of integrals in a 16-dimensional space although many of
the variables enters the functions $\hat{U}$,$\hat{u}_1$ and
$\hat{u}_2$ as sums, see
eqs.(\ref{sommaMessPrimo}--\ref{sommaMessUltimo}), and the actual
number of dimensions can be reduced to 7. In this particular case
other tricks can be used to reduce the number of integrations to 5,
but in general we expect that finding the actual solutions of the
equations will be a very challenging problem.  Second and most
important we see that the message functions $Q(U,u_1,u_2)$ and $q(u)$
enter the equations in an implicit form and in an iterative scheme one
should be able to deconvolve the l.h.s., eq. (\ref{lhs}).  At the RS
level, or 1RSB level on a single sample, this can be done using
Fourier transforms.

Looking at the GBP equations on a single instance eq. (\ref{arghat})
and eqs. (\ref{GBP1},\ref{GBP2},\ref{GBP3}) and at the equations in
the average case for the functions $Q(U,u_1,u_2)$ and $q(u)$
eqs. (\ref{plav1},\ref{lhs},\ref{rhs}) one could be tempted to think
that the functions $q(u)$ and $Q(U,u_1,u_2)$ represent the
distributions over different disorder realizations of the
corresponding GBP messages on a given sample.  As we will show in
sections \ref{sgtran} and \ref{physint}, \emph{this interpretation is
  wrong and misleading}.  It is wrong because it will corresponds to
the assumptions that messages passed from the same region are
uncorrelated and it is misleading because it will lead to the
expectation that, being distributions, they are positive definite,
which turns out to be false.  The fact that the message functions are
not positive definite in general represents another technical
difficulty because it means that they cannot be represented as
populations, a fact that could have simplified the evaluation of the
integral equations.

We note that all these difficulties are absent at the Bethe
approximation level. In this case: i) there are no convolutions in the
equations and ii) the message function itself (and not only the
beliefs, see section \ref{physint}) admits a physical interpretation
as a distribution, consequently it can be represented by a population
and the equations can be solved by a population dynamic algorithm.

In the following we will solve the integral equations in the
zero-field paramagnetic phase, where no convolutions are needed (the
same is not true even in the paramagnetic phase if we consider a
maximal cluster larger than the plaquette or in presence of a field).
Then we will take the first steps into the spin-glass phase, studying
the location of the critical temperature $T_c$ and finding the message
functions at temperatures slightly below $T_c$.  These studies pave
the way for an analysis deep in the spin-glass phase that is left for
future work.

\section{The paramagnetic phase}
\label{paramagnetic}

\subsection{The square lattice}

In the high-temperature region we expect the system to be in a RS
paramagnetic phase.  We also expect that CVM is correct at least at
high enough temperature because it amounts to neglect spatial
correlations beyond a fixed length while the actual correlation length
goes to zero at infinite temperature making the approximation more
precise at higher temperature.  In the paramagnetic phase the symmetry
breaking fields $u$ vanish meaning that the spins have no local
magnetization.  Thus in this region the variational equations admit a
solution of the following kind:
\begin{eqnarray}
q(u) & = & \delta (u) \\
Q(U,u_1,u_2) & = & Q(U)\delta(u_1)\delta(u_2)
\end{eqnarray}
where $Q(U)$ satisfies the following self-consistency equation
\begin{multline}
Q(U) = \int \delta\left[U - \arctanh\big(\tanh(\beta(U_L+J_L))
\tanh(\beta(U_U+J_U)) \tanh(\beta(U_R+J_R)) \big)/\beta\right]\\
dP(J_L) dP(J_U) dP(J_R) dQ(U_L) dQ(U_U) dQ(U_R)
\label{PARAQU}
\end{multline}
The corresponding {\it variational} free energy is given by the
following expression
\begin{multline}
- \beta F = \ln(2) - 2 \int \ln\Big[\cosh\big(\beta(J+U_1+U_2)\big)\Big]
dP(J) dQ(U_1) dQ(U_2)
+ 4 \int \ln\Big[\cosh\big(\beta(J+U)\big)\Big] dP(J) dQ(U)\\
+ \int \ln\Big[1+\tanh\big(\beta(J_L+U_L)\big)
\tanh\big(\beta(J_U+U_U)\big)\tanh\big(\beta(J_R+U_R)\big)
\tanh\big(\beta(J_D+U_D)\big)\Big]\\
dP(J_L) dP(J_U) dP(J_R) dP(J_D) dQ(U_L) dQ(U_U) dQ(U_R) dQ(U_D)
\end{multline}
As we can see the fact that $u$'s vanish simplify considerably the
problem because we do not have convolutions to take and the functions
$Q(U)$ (that in this case \emph{can} be interpreted as a probability
distribution) can be obtained either through a population dynamics
algorithm or by solving iteratively a discretized version of the
variational equation. For symmetry reason the solution is such that
$Q(U)=Q(-U)$.  In spite of its relative simplicity the paramagnetic
solution in the plaquette approximation yields very interesting
results.

We start considering the 2D Edwards-Anderson model with bimodal
interactions $J=\pm 1$.  Although an analytical solution of the 2D
Edwards-Anderson model is missing, numerical studies indicate that the
critical temperature of the model is likely to be zero.  Moreover a
very recent analytical study by Ohzeki and Nishimori \cite{Nishi}
finds strong evidences for the absence of a finite-temperature spin
glass transition in 2D spin glass models.  In the following section we
will study the possibility of a second order spin-glass phase
transition looking for a temperature where small non-zero fields $u$
develops.

Before entering into the details we summarize the main results:
\begin{itemize}
\item the paramagnetic phase is thermodynamically stable down to zero
  temperature, in the sense that it predicts always a positive
  entropy.
\item $Q(U)$ converges to a distribution concentrated on the integers
  in the zero temperature limit yielding a positive zero-temperature
  entropy.
\item there is no spurious spin-glass phase transition (see next
  section).
\end{itemize}

\begin{figure}[htb]
\begin{center}
\epsfig{file=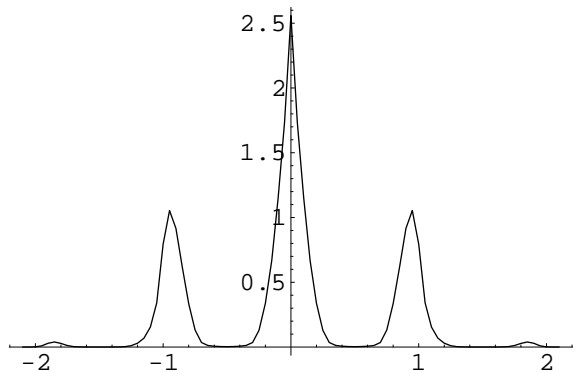,width=8cm}
\caption{The message function $Q(U)$ of the paramagnetic RS plaquette
  approximation at $T=.1$, its support converge on the integers in the
  $T \rightarrow 0$ limit.}
\label{figQU}
\end{center}\end{figure}

In fig. \ref{figQU} we show $Q(U)$ at temperature $T=.1$, it is
already clear that the solution is converging over the integers. At
$T=0$ the population converges to a distribution concentrated over
integers values even if the starting point was a distribution
concentrated on real values, in particular we have: $Q(0)=.534$,
$Q(1)=Q(-1)=.226$, $Q(2)=Q(-2)=.006$, $Q(3)=Q(-3)=O(10^{-6})$.

In the case of bimodal distribution of the couplings the fact that the
distribution concentrates over integer values is usually considered an
indication on the quality of the approximation, we will discuss this
issue in more depth below when studying the behaviour of the specific
heat.  We anticipate that in 3D the function $Q(U)$ does not converge
over the integers at low temperatures, thus indicating that the
paramagnetic solution is not good down to zero temperature in
agreement with the expectation that there is a finite temperature
spin-glass phase transition in $3D$.

Furthermore convergence on the integers is necessary in order to
recover the expected high degeneracy of the ground state energy and
correspondingly a non-zero entropy at zero temperature.  Indeed the
zero temperature entropy can be computed either studying the behaviour
of the free energy at low temperatures or working directly at zero
temperature. The latter approach usually yields more precise estimates
and we have followed it to compute the zero-temperature entropies
reported below. To work at zero temperature one needs to consider the
so-called evanescent fields writing $U=k+\epsilon T$, where $k$ is an
integer and $\epsilon$ is a real number describing the deviations from
the integers at small finite temperature.  Then the function $Q(U)$ is
replaced by a function $Q(k,\epsilon)$ and both the zero temperature
energy and entropy can be expressed in terms of $Q(k,\epsilon)$.  When
there is no convergence over the integers one can in principle study
the solution that is obtained considering only integers values (the
equations are stable on the integers). In this case however the lack
of convergence problems shows up when one tries to compute the zero
temperature entropy, because the evanescent fields $\epsilon$ diverges
and have no stable distribution.

\begin{figure}[htb]
\begin{center}
\epsfig{file=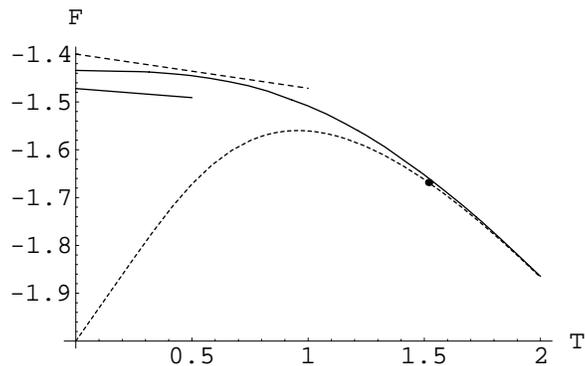,width=8cm}
\caption{Free energy vs. Temperature of the paramagnetic solution in
  the plaquette approximation (solid) for the 2DEA model with bimodal
  couplings. The paramagnetic Bethe solution (dotted) is unstable
  below $T=1.5186$ (dot), the model on the Bethe lattice has a
  spin-glass phase transition at this temperature. The straight lines
  are $E_0-T S_0$ where $E_0$ ($S_0$) is the ground state energy
  (entropy) for the true model (dashed) and for the Bethe lattice
  (solid) from numerics.}
\label{figFTreg}
\end{center}\end{figure}

In fig. \ref{figFTreg} we plot the CVM free energy as a function of
the temperature.  At zero temperature we find $E_0=-1.43404$ that has
to be compared with the best numerical estimate $E_0=-1.401938(2)$
\cite{PA,CHK}. The Bethe approximation result is instead
$E_0=-1.472(1)$ and $S_0=0.0381(15)$ (from a numerical study on the
Bethe lattice \cite{Boett}).  Note that the Bethe approximations
predicts also a spurious spin-glass phase transition at a temperature
$T=1.51865$.  Thus we see that the estimate of the ground state energy
is better than that of the Bethe approximation. Nevertheless the
estimated zero temperature entropy $S_0=0.010(1)$ is too low, compared
to the expected value $S_0=0.0714(2) $ \cite{BGP,LGMM}, the reason for
this is evident from the figure: the quality of the CVM approximation
decreases approaching $T=0$ where the correlation length of the actual
model is expected to diverge.

We recall that the Bethe approximation has the peculiar property that
it is correct on the Bethe lattice.  Therefore there always exists a
thermodynamically stable solution in the Bethe approximation, {\it
  i.e.} the one that yields the free energy of the Bethe lattice.
This is not true for a general CVM approximation and it is the
so-called {\it realizability} problem in CVM theory
\cite{PEL}. However it was noted in \cite{PELrea} that on some models
the plaquette approximation yields the exact result for the free
energy.

\begin{figure}[htb]
\begin{center}
\epsfig{file=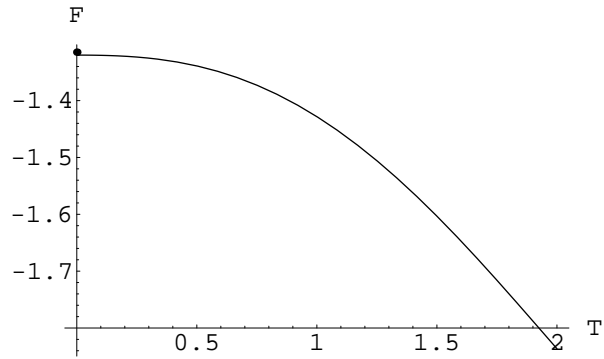,width=8cm}
\caption{Free energy vs. Temperature of the paramagnetic solution in
  the plaquette approximation (solid) for the 2DEA model with Gaussian
  couplings. The dot at $T=0$ corresponds to the value of the ground
  state energy of the actual model.}
\label{figFTgauss}
\end{center}\end{figure}

{\it Gaussian Distributed Couplings} - We considered also the 2DEA
model with Gaussian distribution of the couplings. Again we find that
the paramagnetic phase is thermodynamically stable down to zero
temperature where it predicts a vanishing entropy according to what is
expected.  In this case the CVM estimates are even better than in the
previous case. In fig. \ref{figFTgauss} we plot the free energy as a
function of the temperature, the ground state energy reads
$E_0=-1.3210(2)$ to be compared with the numerical prediction
$E_0=-1.31479(2)$ \cite{CHK}.

\subsection{Triangular and Hexagonal Lattices}

\begin{figure}[htb]
\begin{center}
\epsfig{file=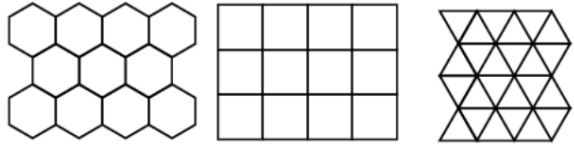,width=8cm}
\caption{We have considered the spin-glass model with bimodal
  couplings on the three regular $2D$ lattices using respectively the
  hexagon, square and triangle approximation}
\label{figlatt}
\end{center}\end{figure}

We studied the spin-glass with bimodal distribution of the couplings
defined on the triangular lattice and on the hexagonal
(a.k.a. honeycomb, brickwork) lattice, using respectively the triangle
and the hexagon as the basic plaquette, see fig. \ref{figlatt}.  Much
as in the square lattice case, the messages are parametrized by a
single function $Q(U)$ in the RS paramagnetic phase representing
respectively the triangle-to-couple and hexagon-to-couple messages.
In the average case, self-consistency equations for the messages are
exactly as Eq.~\ref{PARAQU}, with the only difference being in the
number of hyperbolic tangents contained in the argument of $\arctanh$;
in other words messages satisfy the following equations in
distribution sense
\begin{eqnarray}
U & \eqd & \arctanh\big(\tanh(\beta(U_1+J_1))
\tanh(\beta(U_2+J_2))\big)/\beta\\
U & \eqd & \arctanh\big(\tanh(\beta(U_1+J_1))
\tanh(\beta(U_2+J_2)) \tanh(\beta(U_3+J_3))
\tanh(\beta(U_4+J_4)) \tanh(\beta(U_5+J_5)) \big)/\beta
\end{eqnarray}
for the triangular and hexagonal lattice, respectively.  In both cases
we found again that the paramagnetic phase is thermodynamically
consistent down to zero temperature in the sense that the entropy of
the paramagnetic solution is always positive.  The function $Q(U)$
converges on the integers for the triangular and on the half-integers
for the honeycomb lattice predicting a non-vanishing entropy at zero
temperature in both cases.

\begin{figure}[htb]
\begin{center}
\epsfig{file=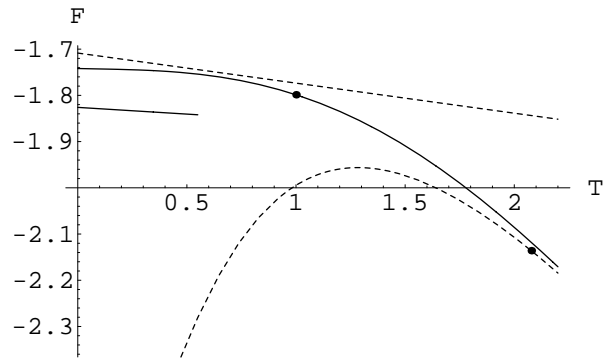,width=8cm}
\caption{Free energy vs. Temperature of the paramagnetic solution in
  the triangle approximation (solid) for the 2DEA model with bimodal
  couplings on the triangular lattice. The entropy is positive down to
  zero temperature, however the dot on the curve marks the temperature
  where a spin-glass solution should be found. The paramagnetic Bethe
  solution (dotted) is unstable below $T=2.0780869$ (dot), the model
  on the Bethe lattice has a spin-glass phase transition at this
  temperature. The straight lines are $E_0-T S_0$ where $E_0$ ($S_0$)
  is the ground state energy (entropy) for the true model (dashed)
  \cite{Poulter2001} and for the Bethe lattice (solid) from numerics
  \cite{Boett}.}
\label{figFTTRI}
\end{center}\end{figure}

In fig.~\ref{figFTTRI} we plot the free energy of the triangular
lattice as a function of the temperature. The most accurate
predictions for the ground state energy and entropy come from a
Pfaffian method \cite{Poulter2001}, giving $E_0= -1.7085(1)$ and $S_0
= 0.065035(2)$.  The Bethe approximation predictions \cite{Boett} are
$E_0=-1.826(1)$ and $S_0=0.0291(10)$ while the present CVM triangle
predictions are $E_0=-1.74227$ and $S_0=0.0087(7)$.  We see that much
as in the above cases the CVM estimate of the free energy largely
improves upon the Bethe one.  Note that the Bethe approximation
predicts a spurious low-temperature spin-glass phase that appears at
$T=2.078086$.  The analysis of the next section shows that the
triangle approximation predicts a spurious spin-glass phase transition
at $T=1.0$.  A detailed study of the spin-glass solution goes beyond
the scope of this work but we expect that it will improve the estimate
of the ground state energy and entropy.  As in the square lattice
case, the zero temperature entropy is less precise than the energy for
the same reasons discussed above.

\begin{figure}[htb]
\begin{center}
\epsfig{file=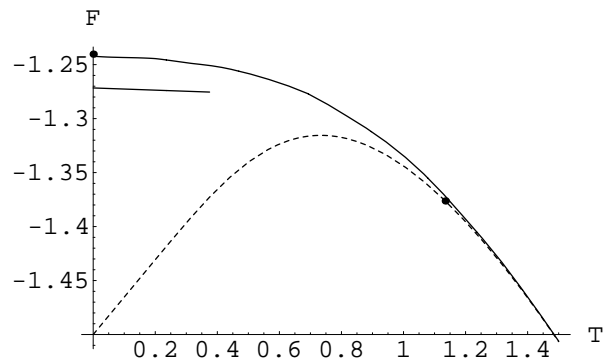,width=8cm}
\caption{Free energy vs. Temperature of the paramagnetic solution in
  the hexagon approximation (solid) for the 2DEA model with bimodal
  couplings on the Hexagonal lattice. The dot at $T=0$ represents the
  ground state energy of the actual model. The paramagnetic Bethe
  solution (dotted) is unstable below $T=1.13459$ (dot), the model on
  the Bethe lattice has a spin-glass phase transition at this
  temperature. The straight line is $E_0-T S_0$ where $E_0$ ($S_0$) is
  numerical estimate for the ground state energy (entropy) for the
  Bethe lattice (solid).}
\label{figFTHEX}
\end{center}\end{figure}

In fig. \ref{figFTHEX} we plot the free energy of the hexagonal
lattice as a function of the temperature. The numerical predictions
for the ground state energy and entropy are respectively
$E_0=-1.2403(2)$ and $S_0=0.02827(5)$ \cite{Arom}.  The corresponding
Bethe lattice predictions \cite{Boett} are $E_0=-1.2716(1)$ and
$S_0=0.0102(10)$ while the present hexagonal CVM predicts
$E_0=-1.242187$ and $S_0=0.020$. In this case also the zero
temperature entropy improves over the Bethe result.  In the case of
the hexagonal lattice both the Bethe and CVM approximations are much
more precise than on the square lattice, and the CVM approximation
corrects $95\%$ of the error of the Bethe approximation.  Note that
the Bethe approximation predicts once more a spurious low temperature
spin-glass phase that appears at $T=1.13459$.  We have not
investigated the presence of a spurious low temperature spin glass
phase in the hexagonal CVM approximation, since this study is
computationally heavy and we strongly expect such a phase not to
exist. Indeed the hexagonal CVM approximation is more accurate than
the square CVM approximation, and the latter does not shows any
spurious spin-glass phase at low temperatures.

In the 2D square lattice spin-glass models we have found that the
paramagnetic phase is thermodynamically stable down to zero
temperature. This is in agreement with numerical evidence that show
that the only critical point is at $T=0$.  Therefore it is natural to
ask whether the CVM approximation also predicts a zero-temperature
critical point.  One can argue that this is not the case. Indeed both
for models with Gaussian and bimodal interactions it can be seen that
a magnetic field scaling with the temperature as $H=h T$ with $h$
small has an effect $O(T)$ on the free energy, therefore the
derivatives with respect to the field diverge as
$d^kF/dH^k=O(T^{1-k})$. On the other hand the
fluctuation-Dissipation-Theorem tells us that the quartic derivative
is related to the fluctuations of the overlap ({\it i.e.}\ the
spin-glass susceptibility) times $T^{-3}$. Thus we see that a quartic
derivative diverging as $T^{-3}$ does not imply a divergent spin-glass
susceptibility at $T=0$.

\subsection{Free energy fluctuations}

In random systems the free energy fluctuates from sample to sample.
The scaling of the variance with the system size is rather nontrivial
in mean-field spin-glass models like the Sherrington-Kirkpatrick model
\cite{CPSV,PR2,ABMM} and in random graphs with fixed connectivity with
bimodal interactions \cite{BKM,PRdil}.  On the other hand in finite
dimensional systems it is known that the variance must scale as the
square root of the volume \cite{MATH,BKM}.  In particular at system
size $N$ the fluctuations $\Delta f_J \equiv f_J-\langle f_J \rangle$
of the free energy $f_J$ of a given sample around its average value
obey:
\begin{equation}
\langle \Delta f_J^2 \rangle={\sigma^2 \over N}
\end{equation}
Although the mean-field prediction is $\sigma=0$ it has been shown
that the loop corrections leads to a non-zero $\sigma$
\cite{TAMAS,AM}. The present approach also predicts a non-zero
$\sigma$ and it allows to get a quantitative estimate.

In the replica framework it can be shown that $\sigma$ is related to
the $O(n^2)$ term in the expansion of $n \Phi(n)$ at small $n$ (see
\cite{CPSV}). It has been recently noted \cite{PRdil} that on the
Bethe lattice one can compute the $O(n^2)$ expanding the variational
expression of $n \Phi(n)$ around $n=0$. Since the expression is
variational the total second derivative with respect to $n$ coincides
with the partial derivatives evaluated at $n=0$.  The same argument
applies in any CVM approximation. In order to obtain the results we
need the variational expression of the CVM free energy written in
terms of the message functions that it is presented in full generality
in appendix \ref{GSP}, the following discussion is based therefore on
the definitions and results of appendix \ref{RSBparametrization} and
\ref{GSP}.  One has to expand the expression of the variational free
energy eq. (\ref{e38}) in powers of $n$ at the second order and
evaluate it using the $Q(U)$ corresponding to $n=0$. We immediately
see that the $O(n^2)$ is given by the sum over the different regions
$r$ of the fluctuations of the corresponding free energy with the
usual region coefficients $c_r$.  In particular we define the
free-energy variance of region $r$ at generic number of RSB steps $K$
as:
\begin{equation}
\sigma^2_r \equiv \langle \langle (\Delta F_{J,r}^{(K-1)})^2\rangle
\rangle-\langle \langle \Delta F_{J,r}^{(K-1)}\rangle \rangle^2
\end{equation}
where we have used the definitions:
\begin{equation}
\langle \langle O \rangle \rangle=\int \prod_{r''s'' \in M(r)}
P_{r''s''}^{(K)}dP_{r''s''}^{(K-1)}\langle O \rangle_J 
\end{equation}
We recall that the RS approximation corresponds to $K=0$. With the
previous definitions we have:
\begin{equation}
\sigma^2={1 \over N}\sum_{r \in R} c_r\sigma_r^2
\end{equation}  
As usual if we assume that the distribution of the couplings $P(J)$ is
the same for all $J_{ij}$ we can conclude that the contributions of
regions with the same form are equal and replace the above expression
with a sum over types of region ({\it e.g.}\ the plaquette, the couple
and the point in the plaquette approximation) each multiplied by the
number of regions of a given type per spin, see discussion below
eq. (\ref{e38}). Note that the $c_r$ can be negative and therefore a
wrong approximation could yield a negative $\sigma^2$, for instance a
negative $\sigma^2$ is predicted by the RS solution of the SK model
below the critical temperature \cite{PR2}, in agreement with the fact
that the correct solution is not RS.  On the contrary in all the $2D$
models considered we have found that the paramagnetic RS solution of
the CVM approximation yields a positive $\sigma^2$ down to zero
temperature in agreement with the expectation that the actual model is
paramagnetic at any finite temperature.

For completeness we report the expression of the free energy
fluctuations $\sigma_{square}^2$ of the plaquette approximations for
the square lattice 2D model in the paramagnetic RS approximation.
According to the above equations we have:
\begin{equation}
\sigma_{square}^2=\sigma_{plaquette}^2-2 \sigma_{couple}^2+\sigma_{point}^2
\end{equation}
In the paramagnetic approximation all the symmetry-breaking small
fields vanish, {\it i.e.} $q(u)=\delta(u)$ and
$Q(U,u_1,u_2)=Q(U)\delta(u_1)\delta(u_2)$ therefore
$\sigma_{point}^2=0$. For the couple we have:
\begin{equation}
\Delta F^{(-1)}_{couple}(U_U,U_D,J) \equiv {1 \over \beta}\ln \cosh \beta (U_U+U_D+J)
\end{equation}
where we have neglected unimportant constant factors. The contribution
of the couple reads:
\begin{equation}
\sigma_{couple}^2 = \int {1 \over \beta^2} \ln^2 \cosh \beta
(U_U+U_D+J) dQ(U_U)dQ(U_D)dP(J)- \left({1 \over \beta} \int \ln \cosh
\beta (U_U+U_D+J) dQ(U_U)dQ(U_D)dP(J) \right)^2
\end{equation} 
For the plaquette we have:
\begin{equation}
\Delta F^{(-1)}_{plaquette}(\# ) \equiv {1 \over \beta}\ln
\sum_{\sigma_1,\sigma_2,\sigma_3,\sigma_4}
\exp\beta((U_{D}+J_{D})\sigma_1 \sigma_2+(U_{L}+J_{L})\sigma_2
\sigma_3+(U_{U}+J_{U})\sigma_3 \sigma_4+(U_{R}+J_{R})\sigma_1
\sigma_4)
\end{equation}
where the argument $\#$ stands for
$(U_{D},U_{L},U_{U},U_{R},J_{D},J_{L},J_{U},J_{R})$ and:
\begin{eqnarray}
\sigma_{plaquette}^2 & = &  \int \Delta F^{(-1)}_{plaquette}(\# )^2
dQ(U_{D})dQ(U_{L})dQ(U_{U})dQ(U_{R})dP(J_{D})dP(J_{L})dP(J_{U})dP(J_{R})
\nonumber 
\\
& - & \left( \int \Delta F^{(-1)}_{plaquette}(\# )
dQ(U_{D})dQ(U_{L})dQ(U_{U})dQ(U_{R})dP(J_{D})dP(J_{L})dP(J_{U})dP(J_{R})
\right)^2 
\end{eqnarray}

In particular at zero temperature we have $\sigma_{square}^2=0.1678$,
$\sigma_{hex}^2=0.100$ and $\sigma_{tri}^2=0.373$ respectively for the
square, hexagon and triangle CVM approximation of the corresponding
$2D$ lattices with bimodal coupling.  In the case of the square
lattice with Gaussian couplings we estimate $\sigma^2=.536$ leading to
$\sigma=.732$ in very good agreement with the value $\sigma=.725$ for
the actual model reported in \cite{BKM}.

We note that the estimates for the $2D$ lattices with bimodal
interactions represent a critical improvement with respect to the
Bethe lattice where one has $\sigma=0$ because of the spatial
homogeneity of the model \cite{BKM,PRdil}.

\subsection{Specific heat at $T=0$}

Another interesting prediction of the CVM approach regards the
behaviour of the specific-heat at low temperatures.  In the case of
the square lattice it was suggested long ago by Swendsen and Wang
\cite{CV1} that the behaviour of the specific heat at low temperature
is of the form
\begin{equation}
c_V \approx  {1 \over T^p}a\exp[-A/T]
\label{cV}
\end{equation}  
with $A=2$. This is absolutely non-trivial because the energy of any
finite-size excitation for the square lattice is a multiple of $4$,
and this would lead instead to $A=4$ as later claimed in \cite{CV5}.
Over the years the $A=2$ result has been supported by many authors
\cite{CV2,CV3,LGMM,CV4}; recently another scenario has also been
proposed in which $c_V$ behaves as a power law \cite{CV6} with a
universal exponent that is the same of models with Gaussian
distributions of the couplings. The true nature of $c_V$ remains
nevertheless unclear \cite{CV7}.

Given that $Q(U)$ is symmetric it follows that in both the square,
hexagonal (in agreement with \cite{PoulterAtisattapong}) and
triangular lattice the behaviour of the specific heat at low
temperature predicted by the CVM approximation is of the form
(\ref{cV}) with $A=2$ and $p=2$.  Note that this is a non-trivial
prediction not only for the square lattice but also for the triangular
one.  It must be remembered however that the CVM is intrinsically a
mean-field approximation and could never give a power-law behaviour.
On the other hand it clearly suggests that even if a power-law
behavior is actually present there are corrections of the form
(\ref{cV}) with the non-trivial value $A=2$.

The CVM approach yields also the numerical coefficient $a$ of the
leading term in (\ref{cV}) that can be computed working directly at
zero temperature.  Given that the correction is exceedingly small it
is safer to work directly at zero temperature.  In order to compute
this coefficient one has to write the finite-temperature field $U$ as
\begin{equation}
U=k+T \epsilon + T e^{-2 \beta} z
\end{equation}
and study the joint distribution of the triplets $Q(U) \longrightarrow
Q(k,\epsilon,z)$ whose equation was obtained considering the leading
order contribution of the equation of $Q(U)$.  Summarizing the low
temperature behaviour of the specific heat according to the CVM
approximation for the 2D Ising spin-glass with bimodal interactions is
\begin{equation}
c_V={a \over T^2}\exp[-2/T]
\end{equation}
The coefficient in the case of the square lattice is $a \approx 60$
and does not appear to compare well with the numerical data, if we go
back to fig. \ref{figFTreg} we can argue that the error of the CVM
square approximation with respect to the actual model is not small
enough to reproduce quantitatively the behaviour of the specific heat

At last, it is also interesting to note that much as in the case of
$O(n^2)$ term discussed above, the replica CVM predicts a {\it
  qualitatively} different behaviour than the Bethe approximation. The
low-temperature specific-heat behaviour has not been considered in the
case of the Bethe lattice; nevertheless it is known that the Bethe
approximation yields a spurious phase transition with non-zero
symmetry-breaking fields $u$ concentrated over the integers. In the
case of odd connectivities (corresponding {\it e.g.}\ to the hexagonal
lattice) this leads naturally to a $A=2$ gap, instead in the case of
even connectivities (corresponding to the square and triangular
lattices) the fields are known to concentrate over {\it odd} integers
values leading to $A=4$, while as we saw the use of the CVM
approximation reduces the gap and leads to $A=2$ for all the three 2D
model considered.

\section{The spin-glass phase transition in the plaquette approximation}
\label{sgtran}

A typical application of the CVM \cite{CVM} is the location of the
critical temperature of phase transitions.  We recall that in the
plaquette approximation above the critical temperature the small
fields $u$ vanish, {\it i.e.}\ we have:
\begin{eqnarray}
q(u) & = & \delta (u)\\
Q(U,u_1,u_2) & = & Q(U)\delta(u_1)\delta(u_2)
\end{eqnarray}
A spin-glass phase transition corresponds to the fact that the
symmetry-breaking fields $u$ become non-zero.  Near the critical
temperature of a second-order phase transition the symmetry-breaking
fields $u$ will be no longer zero but small, and we will determine the
location of the critical temperature considering the second moments of
the distributions.  We define:
\begin{eqnarray}
a & := & \int q(u) u^2 du
\label{eqA}
\\
a_0(U) & := & \int \int Q(U,u_1,u_2) du_1 du_2 
\\
a_{ij}(U) & :=  & \int Q(U,u_1,u_2)u_i u_j  du_1 du_2    \ \ \ i,j=1,2
\label{eqA1122}
\end{eqnarray} 
For symmetry reason we expect
$Q(U,u_1,u_2)=Q(-U,-u_1,u_2)=Q(U,-u_1,-u_2)=Q(U,u_2,u_1)$ thus
$a_{11}(U)=a_{22}(U)$, furthermore $a_0(U)$ and $a_{11}(U)=a_{22}(U)$
will be even function of $U$, while $a_{12}(U)$ will be an odd
function of $U$.

Now to explain the basics of the method we consider the simplest case
of the Bethe approximation in which only the function $q(u)$ is
present.  In this case the transition is marked by the fact that the
parameter $a$ defined above vanishes above the critical temperature
while it is different from zero below.  To determine the critical
temperature one expands the iterative equation at small $a$ and obtain
something of the form:
\begin{equation}
C(T)a+B(T)a^2+O(a^3)=0 \ ,
\label{JacBethe}
\end{equation}
the critical temperature corresponds to the vanishing of the
coefficient $C(T)$, or equivalently to the fact that the homogeneous
linear equation $a C(T)=0$ admits a non-zero solution.  The condition
$C(T)=0$ leads to the equation $\langle \tanh^2 \beta_c J
\rangle_J=1/c$ in the Bethe approximation where $c+1$ is the
connectivity of the model.  We note that below the critical
temperature the function $q(u)$ is also described by higher order
moments, however to determine exactly the critical temperature it is
sufficient to consider the behavior of the second moment $a$.

We have obtained the corresponding linear homogeneous equation for the
variables $\{a, a_{11}(U),a_{12}(U)\}$ in the form:
\begin{eqnarray}
a & = & K_{a,a} a+ \int dU' K_{a,a_{11}}(U')a_{11}(U')+ \int dU'
K_{a,a_{12}}(U')a_{12}(U')
\label{nuc1}
\\
a \, a_0(U)+a_{11}(U) & = & K_{a_{11},a}(U) a+ \int dU'
K_{a_{11},a_{11}}(U,U')a_{11}(U')+ \int dU'
K_{a_{11},a_{12}}(U,U')a_{12}(U') 
\label{nuc2}
\\
a_{12}(U) & = & K_{a_{12},a}(U) a+ \int dU'
K_{a_{12},a_{11}}(U,U')a_{11}(U')+ \int dU'
K_{a_{12},a_{12}}(U,U')a_{12}(U') 
\label{nuc3}
\end{eqnarray}
where the various coefficients $K$ depends on the temperature and on
the corresponding function $Q(U)$, we do not report them all but in
the following we will explain how they have to be obtained.  The
critical temperature should be identified with the point where the
above homogeneous set of equations admits a non-zero solution for the
parameters $(a,a_{11}(U),a_{12}(U))$. The corresponding eigenvector
determines the behavior of the function $q(u)$ and $Q(U,u_1,u_2)$
slightly below the critical temperature.

To see how the coefficients $K$ have to be obtained we consider the
equation for the parameter $a$.  We start from the equation for
$q(u)$, eq. (\ref{arghat}), we multiply both sides times $u^2$ and
integrate over $u$.  We have:
\begin{equation}
a= \int \hat{h}^2dQ_{bcfg,fg}
dQ_{fglm,fg}dq_{cg,g}dq_{gh,g}dq_{gm,g}dP(J_{fg}) 
\label{a}
\end{equation}
Where the function $\hat{h}$ is defined in eq. (\ref{arghat}).  Now we
expand the function $\hat{h}$ at the second order in powers of the
small fields $u$, and express the integrals in eq. (\ref{a}) in terms
of $a$, $a_0(U)$ and $a_{ij}(U)$, the result is:
\begin{eqnarray}
a & = & 3 a \int P(J)a_0(U^P)a_0(U^D)\tanh^2 \beta
(J+U^D+U^P)\,dJ\,dU^D\, dU^P+ 2 \int a_{11}(U) dU+ 
\\
&+ & 2 \int P(J)a_0(U^P)a_{11}(U^D)\tanh^2 \beta
(J+U^D+U^P)\,dJ\,dU^D\, dU^P+ 
\\
& + & 4  \int P(J)a_0(U^P)a_{12}(U^D)\tanh \beta
(J+U^D+U^P)\,dJ\,dU^D\, dU^P 
\end{eqnarray}
The above equation corresponds to eq. (\ref{nuc1}).  The other
coefficients can be obtained similarly multiplying both sides of the
equation for $Q(U,u_1,u_2)$ ({\it i.e.}\ eqs. (\ref{lhs}) and
(\ref{rhs})) times $(\tilde{u}_{gflm,lm}^l)^2$ and
$\tilde{u}_{gflm,lm}^l \tilde{u}_{gflm,lm}^m$ and integrating
over. The resulting expressions are fairly complicated and we do not
write them down here.

A complete analytical treatment of the problem seems unfeasible, also
because we do not have an analytical expression of $Q(U)$ at all
temperatures.  Thus we have discretized the space of the $U$, assuming
that it can takes only a finite number $2 I_{max}+1$ of values in the
interval $(-U_{max},U_{max})$, {\it i.e.}\ $U=i\, du$ with $i=0, \pm
1,\pm 2, \dots ,\pm I_{max}$ and $du=U_{max}/I_{max}$. Correspondingly
we have a set of $4 I_{max}+3$ variables ${\bf a} \equiv
(a,a_{11}(U),a_{12}(U))$.  At any temperature we first compute $Q(U)$
on the discretized range of $U$ and then we rewrite the set of
equation (\ref{nuc1},\ref{nuc2},\ref{nuc3}) in the form ${\bf J} \cdot
{\bf a}=0$ where ${\bf J}$ is a $(4 I_{max}+3) \times (4 I_{max}+3)$
matrix, (in the following we will refer to it as the Jacobian matrix).

The computation of the coefficients is the technical bottleneck of the
computation, we have worked typically with $U_{max}=2$ and
$I_{max}=40$. The Jacobian matrix is diagonalized and the critical
temperature have to be identified with the point where it has a
vanishing eigenvalue or correspondingly a vanishing determinant.  Note
that we are linearizing the equations and therefore we call {\it
  Jacobian} the matrix that we compute, as a consequence this matrix
is not symmetric and can have complex eigenvalues. A symmetric matrix
would be obtained had we considered the Hessian of the free
energy. However as far as the determination of the critical
temperature is concerned the two approaches are completely equivalent.

\begin{figure}[htb]
\begin{center}
\epsfig{file=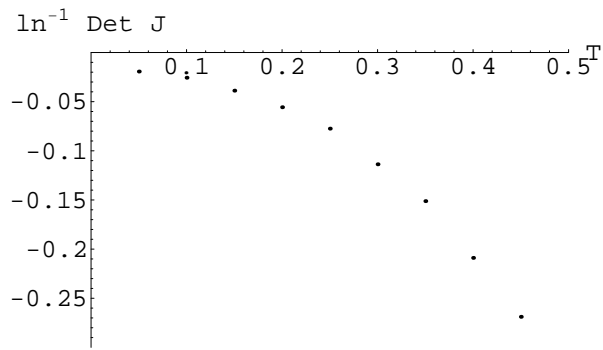,width=8cm}
\caption{Plot of the inverse of the logarithm of the determinant of
  the Jacobian vs. Temperature for the 2D square lattice with bimodal
  interactions (see text). It is strictly positive down to zero
  temperature thus there is no second-order spin-glass phase
  transition in the model at finite temperature. }
\label{figdetJ2D}
\end{center}\end{figure}

We have considered the 2DEA Ising model with couplings $J=\pm 1$ on
the square lattice.  Interestingly enough \emph{the plaquette
  approximation predicts no spurious spin-glass phase transition at
  any finite temperature}. In other words the determinant of the
Jacobian remains always finite, this has to be compared with the Bethe
approximation that yields a spurious spin-glass transition at
$T=1.51865$ for the 2DEA model with $J=\pm 1$.  In
fig. \ref{figdetJ2D} we plot the inverse of the logarithm of the
determinant of the Jacobian at low temperatures. This was obtained
using $U_{max}=2.1$ and $I_{max}=42$ therefore the Jacobian is a $171
\times 171$ matrix. The plot shows that the determinant does not
vanish at least down to $T=.05$ and suggest that it does not vanish at
any finite temperature. A careful study of its behavior as $T
\rightarrow 0$ goes beyond the scope of this work.

A similar study on the 2DEA Ising model with couplings $J=\pm 1$ on
the triangular lattice shows that instead the Jacobian vanishes at
$T=1.0$ that improves considerably on the Bethe lattice estimate
$T=2.078$.  It is interesting to note that also in the 2D triangular
antiferromagnet (which has again a zero temperature critical point)
the CVM approximations yields a spurious phase transition \cite{PP}.

\subsection{General dimension}

The plaquette approximation can be also applied to regular lattices in
any number of dimensions. The objects to be considered are still the
messages functions $q(u)$ and $Q(U,u_1,u_2)$ but the coefficients
$c_r$ of the regions and the total number of regions changes. For
instance in $3D$ we have $c_\text{plaquette}=1$, $c_\text{couple}=-3$
and $c_\text{point}=7$ while the number of regions per spin are
$n_\text{plaquette}=3$, $n_\text{couple}=3$, $n_\text{point}=1$.  The
total number of messages entering in a given region also changes, in
particular in generic dimension $D$ there are $2D$ messages $q(u)$
entering on the point; on the couple of points there are $2D-2$
messages $Q(U,u_1,u_2)$ and $2D-1$ messages $q(u)$ for each point,
while on the plaquette there are $2D-3$ messages $Q(U,u_1,u_2)$ on
each link and $2D-2$ messages $q(u)$ on each point.  The above
formalism for the study of a second-order phase transition can be
extended straightforwardly to general dimension provided some care is
taken in order for the computation of the Jacobian matrix to be done
in reasonable time.  In practice we have introduced auxiliary
functions to represents the convolutions of $Q(U)$ with itself in
order that the integrals needed to compute the elements of the
Jacobian remain three-dimensional as in 2D.

\begin{figure}[htb]
\begin{center}
\epsfig{file=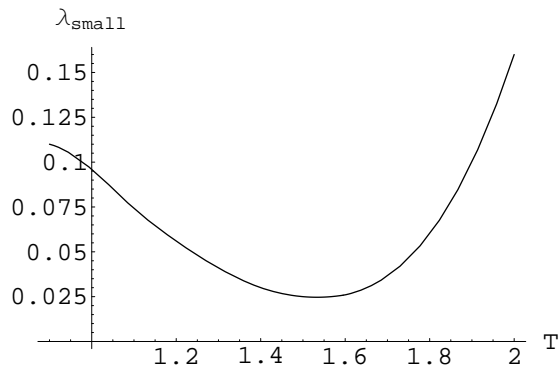,width=8cm}
\caption{Plot of the smallest eigenvalue of the Jacobian around the
  expected critical temperature $T=1.1$ for the plaquette
  approximation of the 3DEA model with bimodal interactions. Since it
  does not vanish the CVM predicts no second-order phase transition in
  this approximation.}
\label{figlam3D}
\end{center}\end{figure}

\begin{figure}[htb]
\begin{center}
\epsfig{file=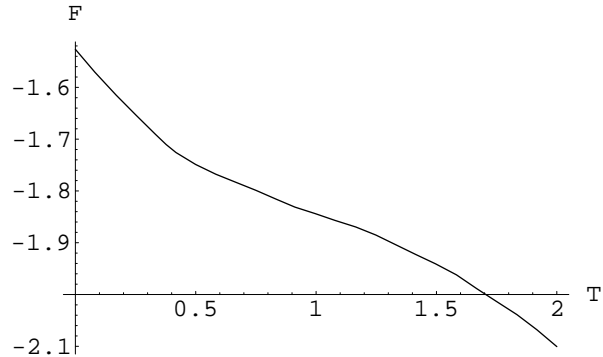,width=8cm}
\caption{Free energy vs. Temperature of the paramagnetic RS solution
  of 3DEA model in the plaquette approximation, although in this
  approximation there is no second-order phase transition, the
  paramagnetic solution is unphysical at low temperature predicting a
  negative specific heat.}
\label{figFT3D}
\end{center}\end{figure}

In dimension higher than two the EA model is largely believed to
display a second-order spin-glass phase transition.  In
fig. \ref{figlam3D} we plot the value of the smallest eigenvalue of
the Jacobian matrix of the 3DEA Ising model. We see that unfortunately
it does not vanish at all, although it decreases considerably around
the temperature where the actual model is believed to have a phase
transition, $T \sim 1.1$. The plaquette approximation leads to a
disaster in 3D: if we did not know the actual behavior of the model we
could wrongly think that much as in 2D the paramagnetic phase is
stable down to zero temperature; however a clear hint that this cannot
be the case comes from the study of the free energy. In
fig. \ref{figFT3D} we plot the free energy as a function of the
temperature, this shows that the entropy remains positive but the free
energy has the wrong convexity at low temperature and negative
specific heat.  Another indication that the paramagnetic solution is
wrong in 3D at low temperature comes from the fact that at zero
temperature the solution does not converge on integers values, at
variance with the 2D case studied in the previous section.  We have
also considered different distributions of the couplings (Gaussian and
Diluted) and check that unfortunately the plaquette approximation
still does not predict any phase transition in $3D$.  The present
approach is able to detect a second-order phase transition where the
variables $(a,a_{11}(U),a_{12}(U))$ are small, it is also possible
that the plaquette approximation makes the transition first-order but
we leave the investigation of this point for future work.  Note that
the smallest eigenvalue gets very near to zero therefore we expect
that in an approximation with a basic region slightly larger than the
plaquette we should be able to recover the expected phase transition.

\begin{figure}[htb]
\begin{center}
\epsfig{file=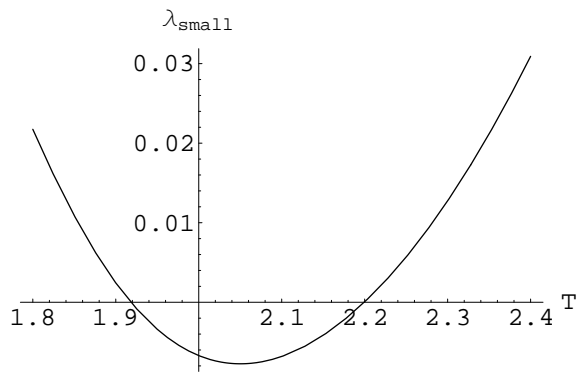,width=8cm}
\caption{The smallest eigenvalue of the Jacobian vs. Temperature for
  the plaquette approximation of the 4DEA model with bimodal
  interactions. It vanishes at $T=2.2$ thus correcting $2/3$ of the
  error of the Bethe approximation $T=2.52$ on the actual value of the
  critical temperature $T=2.03$ \cite{MarinariZuliani1999}. Note the
  presence of second zero at $T \approx 1.9$ that we interprete as a
  sign of RSB, see text.}
\label{figlam4D}
\end{center}\end{figure}

Fortunately enough, the situation gets better in dimension four (see
fig. \ref{figlam4D}) where the smallest eigenvalues vanishes at
$T=2.2$, thus correcting around $2/3$ of the error of the Bethe
estimate $T=2.51$ of the actual value of the critical temperature
$T=2.03$ estimated numerically \cite{MarinariZuliani1999}.  Increasing
the dimension the quality of the results improves systematically, in
$5D$ we have $T=2.550$ to be compared with high-temperature series
estimates $T=2.57(1)$ \cite{HighTemp1} and $T=2.54(3)$
\cite{HighTemp2}, thus correcting almost all the error of the Bethe
approximation estimate $T=2.88$.

It is interesting to observe that according to fig. (\ref{figlam4D})
the smallest eigenvalue in dimension four of the paramagnetic solution
vanishes again lowering the temperature below $T=2.2$ at around
$T=1.9$, we will discuss this unexpected feature of the solution at
the end of the present section.

We have seen that the plaquette approximation gives good results in
$2D$ and in general dimension greater than three, while in $3D$ it
leads to a disaster at low temperature.  We note that the Bethe
approximation is correct at $D=1$ and at $D=\infty$, thus it is
natural that the plaquette approximation is a perturbative correction
to Bethe in high enough dimension as can be seen already in dimension
five.  Looking at the behavior of the smallest eigenvalue of the
Jacobian matrix in $D=3$ we see that, although it does not vanish, it
has a minimum around the true critical temperature $T=1.1$ and it is
likely that the second-order phase transition reappears considering a
basic region larger than the plaquette, {\it e.g.}\ the cube. We note
however that this approximation requires to deal with equations for
the order parameters involving convolutions {\it even} in the RS
paramagnetic phase, a technical difficulty that is not present for the
plaquette.
 
The study of the spin-glass phase requires to deal with non-zero
fields $(u,u_1,u_2)$ and to deal with the convolutions appearing in
eq. (\ref{lhs}). Thus the study of the spin-glass phase requires to
tackle this technical difficulty and goes beyond the scope of this
work. Nevertheless the study of the Jacobian gives us a {\it crucial}
information on the spin-glass phase that we will discuss in the
following.

\begin{figure}[htb]
\begin{center}
\epsfig{file=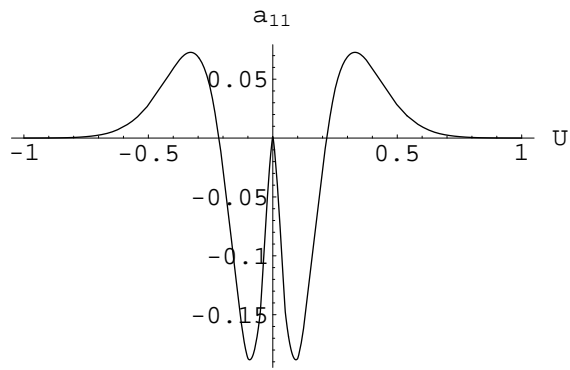,width=8cm}
\caption{The function $a_{11}(U)$ (modulo an unknown positive constant
  scaling as $T_c-T$) at a generic temperature slightly below
  $T_c=2.2$ for the 4DEA with bimodal interactions, see text. It is
  negative for some values of $U$ meaning that the function
  $Q(U,u_1,u_2)$ is not positive definite.}
\label{figa11}
\end{center}
\end{figure}

Slightly below the critical temperature in $D>3$ the quantities
$(a,a_{11}(U),a_{12}(U))$ are proportional to the vanishing
eigenvector of the Jacobian matrix.  In other words we have:
\begin{equation}
(a,a_{11}(U),a_{12}(U))=b(T_c-T)
(\lambda^{(0)}_a,\lambda^{(0)}_{a_{11}}(U),\lambda^{(0)}_{a_{12}}(U)) 
\end{equation}
where
$(\lambda^{(0)}_a,\lambda^{(0)}_{a_{11}}(U),\lambda^{(0)}_{a_{12}}(U))$
are the components of the eigenvector corresponding to the zero
eigenvalue and $b$ is some numerical constant that cannot be
determined solely from the knowledge of the Jacobian but needs the
computation of the quadratic terms.  Indeed the determination of the
proportionality factor requires to include the next order terms
analogous to the term $B(T)$ in eq. (\ref{JacBethe}) at the Bethe
level.  In fig. \ref{figa11} we plot the function $a_{11}(U)$ (modulo
an unknown positive constant scaling as $T_c-T$, {\it i.e.}\ the
non-normalized eigenvector) at a generic temperature slightly below
$T_c=2.2$ for the 4DEA with bimodal interactions, the proportionality
factor is such that the $a$ component is positive, $a=0.08490$.

We see that $a_{11}(U)$ is negative for some values of $U$ and this is
puzzling, indeed we recall that the definition of $a_{11}(U)$ is:
\begin{equation}
a_{11}(U)=\int Q(U,u_1,u_2)\,u_1^2\,du_1\,du_2
\end{equation}
thus if $a_{11}(U)$ is negative below the critical temperature for
some $U$ it follows that \emph{the message function $Q(U,u_1,u_2)$
  cannot be positive definite!}  The first consequence of this fact is
that the function $Q(U,u_1,u_2)$ cannot be interpreted as a
distribution function of the messages $(U,u_1,u_2)$ on a given sample:
had we followed that interpretation we should have concluded that the
whole approach is inconsistent.  In the next section we will discuss
this issue in more depth and see that instead it is the naive
interpretation that is actually inconsistent, in particular we will
show that the message functions $Q(U,u_1,u_2)$ need not to be positive
definite while the beliefs of the regions do.

We mention that negative $a_{11}(U)$ are found also if we study the
response of the system to the presence of a small field $H$ in the
high-temperature phase. In this case we find non-zero values of
$(a,a_{11}(U),a_{12}(U))$ of order $O(H^2)$ that can be determined
inverting the Jacobian matrix and applying it to the $O(H^2)$
perturbation, and again we find that while $a$ is positive $a_{11}(U)$
is negative for some values of $U$.  This effect survives in the
infinite temperature limit. In this regime we find that at leading
order the variables to be considered are $a$ and
$\overline{a}_{11}=\int a_{11}(U)\,dU$ and an explicit computation
shows that $a\simeq H^2 \beta^2$ and $\overline{a}_{11}\simeq -3
\beta^6 H^2 $ in any dimension.

We note that the fact that the messages are not positive definite
means that they cannot be simply represented as populations and this,
together with the presence of convolutions in the variational
equations, is a technical challenge to be overcome in order to obtain
quantitative results for general CVM approximations and for all
regions of the phase diagram.

We also mention that the Jacobian approach presented here can be also
applied to study the phase diagram of models with ferromagnetically
biased interactions, in this case one would be interest in the
location of the ferromagnetic transition and the variables to be used
should be $\tilde{a}=\int q(u) u\, du$ and $\tilde{a}_1(U)=\int
Q(U,u_1,u_2)u_1 du_1du_2 dU$.

We end this section with a comment on the peculiar feature displayed
by the smallest eigenvalue of the Jacobian in four dimension according
to fig. (\ref{figlam4D}). Below $T=2.2$ we expect to find a RS
spin-glass solution with a non-zero positive value of $a \propto
(T_c-T)$. As we already said the actual value of the proportionality
factor cannot be determined solely from the knowledge of the Jacobian
but needs the computation of the quadratic terms. However if we assume
that these terms do not change too much with the temperature between
the $T=2.2$ and $T=1.9$ ({\it i.e.} where the eigenvalue vanishes
again) we easily see that the parameter $a$ should have the opposite
behaviour of the smallest eigenvalue. In particular, lowering the
temperature, it will initially increase from zero to some maximum and
then decrease again to zero. This would be completely unphysical and
means that probably to RS spin-glass solution becomes meaningless and
has to be abandoned below some temperature greater that $T=2.05$ where
the smallest eigenvalue reaches its minimum. It is tempting to
interprete this fact as an indication that the RS spin-glass solution
is physically wrong at low temperature and that a RSB solution has to
be considered instead.  One can further speculate that if this is the
case two phenomena should be observed. First the effect should become
more pronounced while increasing the size of the basic CVM region. In
other words increasing the precision of the CVM approximation the
region of validity of the RS spin-glass solution should shrink to zero
{\it i.e.} the first two zeroes of the smallest eigenvalue should tend
to coincide. Second for a given CVM approximation the use of a RSB
solution should increase the range of validity of the solution
shifting the point where the solution becomes unphysical to lower
temperatures. We think that this is a very interesting open problem.

\section{Physical interpretation of the beliefs}
\label{physint}

In the previous section we have seen that the messages functions
$Q(U,u_1,u_2)$ of the plaquette CVM approximation in general dimension
are not definite positive and cannot be interpreted as distribution
functions.  In this section we will obtain the physical
interpretations of the beliefs of $b_r(\sigma)$ of the regions of the
replicated model. In particular we will show that the beliefs in the
RS approximation can be interpreted as distributions over the disorder
of the local Hamiltonians.

We consider first the belief of a point on the lattice, say $0$. In
the $n$-replicated system there are $n$ spins $\sigma_0^1, \dots ,
\sigma_0^n$ on that point. As we saw in section
\ref{TheReplicaApproach} averaging over the disorder couples the
different replicas and the effective Hamiltonian becomes
$-\sum_{ij}\ln \langle \exp[\beta J
  \sum_a\sigma_i^a\sigma_j^a]\rangle$. The belief $b(\sigma_0)$
describes the marginal distribution of the replicated spins at point
$0$ with respect to the replica Hamiltonian. Any correlation between
any number $p$ of the $n$ spins at site $0$ can be expressed in terms
of $b(\sigma_0)$
\begin{equation}
\langle\langle \sigma_0^{a_1}\dots \sigma_0^{a_p}\rangle\rangle=
\sum_{\sigma_0}(\sigma_0^{a_1}\dots \sigma_0^{a_p})b(\sigma_0) 
\label{corrbel}
\end{equation}
where $\langle\langle \cdots \rangle\rangle$ means average with
respect to the replicated Hamiltonian.  On the other hand such a
correlation can be written as:
\begin{equation}
\langle\langle \sigma_0^{a_1}\dots
\sigma_0^{a_p}\rangle\rangle=\frac{\sum_{\{\sigma\}}(\sigma_0^{a_1}\dots
  \sigma_0^{a_p})\langle e^{\beta
    \sum_{ij,a}J_{ij}\sigma_i^a\sigma_j^a}\rangle}{\sum_{\{\sigma\}}\langle
  e^{\beta \sum_{ij,a}J_{ij}\sigma_i^a\sigma_j^a}\rangle}= \frac{
  \langle \,(\langle \sigma_0^{a_1} \rangle_J \dots \langle
  \sigma_0^{a_p}\rangle_J ) Z_J^n \rangle}{\langle Z_J^n
  \rangle}=\frac{\langle m_{0,J}^p Z_J^n \rangle}{\langle Z_J^n
  \rangle} 
\label{corrbel2}
\end{equation}
Where $\langle \cdots \rangle$ means average over the disorder, $Z_J$
is the partition function of the non-replicated system for a given
realization of the disorder $J$, $\langle \cdots \rangle_J$ means
thermodynamic average at given disorder $J$ and $m_{0,J}$ is the
magnetization at site $0$ of the non-replicated system with a given
disorder realization $J$. The equality between the second and the
third term follows from putting the disorder average outside the sum
over the configurations of the replicated system and thus recovering
the independence of the different replicas prior to the averaging.
The above equations tells us that the correlation between $p$
replicated spins at the same site $0$ is equal to the average with
respect of the disorder of the $p$ moment of the magnetization at site
$0$ of the non-replicated system.  Note that each disorder realization
is weighted with a weight proportional to the partition function to
the power $n$.  In particular when $n \rightarrow 0$ we recover the
standard white average over the disorder while for non-zero $n$ we are
selecting samples with free energy different from the typical one
\cite{PR2}.  According to section \ref{RSBparametrization} the RS
parametrization of the belief $b(\sigma_0)$ is obtained through a
function $p(u)$ (the same for each site) in the form:
\begin{equation}
b(\sigma_0)=\int p(u)\frac{e^{\beta u \sum_a \sigma_0^a}}{(2\cosh
  \beta u)^n}du \ , 
\end{equation} 
now using eq. (\ref{corrbel}) and (\ref{corrbel2}) we finally arrive
at the following equation:
\begin{equation}
\int p(u) (\tanh \beta u)^p\,du=\frac{\langle m_{0,J}^p Z_J^n
  \rangle}{\langle Z_J^n \rangle} 
\label{fund}
\end{equation}
The above equation encodes the physical meaning of the belief function
$p(u)$.  On a given sample $J$ the magnetization $m_{0,J}$ of site $0$
is determined by an effective field $u=\arctanh(m_{0,J})/\beta$ on
site $0$ generated by the rest of the system. Since the above equation
is valid for any $p$ it follows that \emph{the function $p(u)$ is the
  distribution over the different samples of the effective field
  acting on a given site}.

The same interpretation can be obtained for the belief of the couple
of points and for the plaquette.  On a given sample the effect of the
rest of the system on a couple of spins $\sigma_1$ and $\sigma_2$
generates an effective local Hamiltonian of the form $-(U
\sigma_1\sigma_2+u_1 \sigma_1 +u_2 \sigma_2)$ that determines
completely the magnetizations and correlation of the two spin; the
belief function of the couple of points $P(U,u_1,u_2)$ is the
distribution over the different samples of the effective local
Hamiltonian.

\begin{figure}[htb]
\begin{center}
\epsfig{file=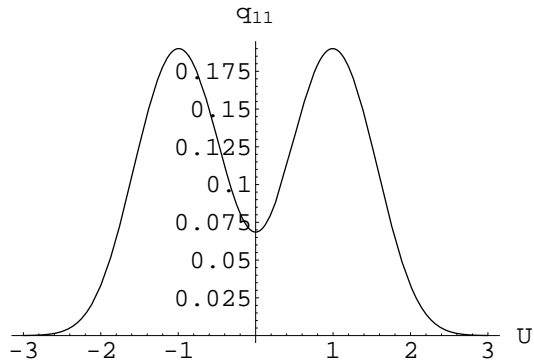,width=8cm}
\caption{The function $q_{11}(U)=\int P(U,u_1,u_2)u_1^2\,du_1\, du_2$
  (modulo an unknown positive constant scaling as $T_c-T$) at a
  generic temperature slightly below $T_c=2.2$ for the 4DEA with
  bimodal interactions, see text. It is always positive as it should
  since $P(U,u_1,u_2)$ is a distribution.}
\label{figq11}
\end{center}\end{figure}

\begin{figure}[htb]
\begin{center}
\epsfig{file=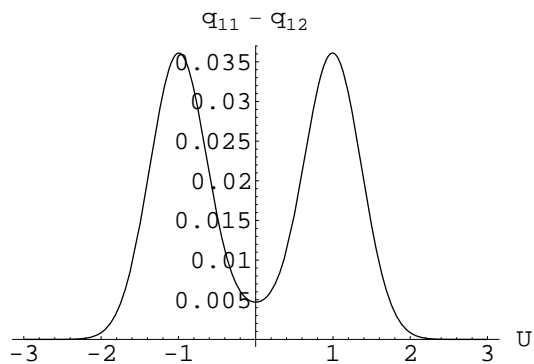,width=8cm}
\caption{The function $q_{11}(U)-q_{12}(U)=\int
  P(U,u_1,u_2)(u_1-u_2)^2/2\,du_1\, du_2$ (modulo an unknown positive
  constant scaling as $T_c-T$) at a generic temperature slightly below
  $T_c=2.2$ for the 4DEA with bimodal interactions, see text. It is
  always positive as it should since $P(U,u_1,u_2)$ is a
  distribution.}
\label{figDQ2}
\end{center}\end{figure}

The CVM approach (in particular eq. (\ref{belief})) tells us that the
distributions $p(u)$ and $P(U,u_1,u_2)$ can be expressed in terms of
the message functions $q(u)$ and $Q(U,u_1,u_2)$ in the following way
(we specialize as before to the 2DEA model on the regular lattice):
\begin{equation}
p(u)=\int \prod_{i=1}^4[q(u_i)du_i]\delta(u-(u_1+u_2+u_3+u_4))
\label{Ppunto}
\end{equation} 
and 
\begin{eqnarray}
P(U,u_1,u_2)& = & \int
\prod_{i=1}^3[q(u_i)du_i]\prod_{i=1}^3[q(v_i)dv_i]
\prod_{a=up,down}[Q(U_a,u_a,v_a)dU_adu_adv_a]P(J)dJ\, 
\times 
\nonumber
\\
 & \times &
 \delta(U-(U_{up}+U_{down}+J))\,\delta\left(u_1-(\sum_{i=1}^3
   u_i+u_{up}+u_{down})\right)\,\delta\left(u_2-(\sum_{i=1}^3
   v_i+v_{up}+v_{down})\right) 
\label{Pcoppia}
\end{eqnarray}
In the Bethe approximation the message function $q(u)$ can be
interpreted as the disorder distribution of the effective field $u$ on
a given spin $\sigma$ with connectivity $c$ when $c-1$ links connected
to it are removed.  This is a peculiar feature of the Bethe
approximation but in general the message functions do not admit an
interpretation as distributions as the one derived above.  Indeed, as
we have seen in the previous section, they are not positive definite
in general.  On the other hand it can be argued that the distribution
$P_{no-J}(U,u_1,u_2)$ of the effective Hamiltonian of a couple of
points in the absence of the link connecting them obeys the equation:
\begin{equation}
P(U,u_1,u_2)=\int dU'dJ P_{no-J}(U',u_1,u_2)\delta(U-(U'+J))P(J)
\end{equation}
and therefore $P_{no-J}(U,u_1,u_2)$ is equal to (\ref{Pcoppia})
without the integration over $P(J)dJ$:
\begin{eqnarray}
P_{no-J}(U,u_1,u_2)& = & \int \prod_{i=1}^3[q(u_i)du_i]
\prod_{i=1}^3[q(v_i)dv_i]\prod_{a=up,down}[Q(U_a,u_a,v_a)dU_adu_adv_a]\,
\times 
\nonumber
\\
 & \times & \delta(U-(U_{up}+U_{down}))\,
 \delta\left(u_1-(\sum_{i=1}^3
   u_i+u_{up}+u_{down})\right)\,\delta\left(u_2-(\sum_{i=1}^3
   v_i+v_{up}+v_{down})\right) 
\label{PnoJ}
\end{eqnarray}
A similar expression can be obtained for the distribution of the
effective Hamiltonian of the plaquette without the links inside it.
We note that the distribution $P_{no-J}(U,u_1,u_2)$ allows a
straightforward computation of the local energy through the following
expression that can be also (consistently) obtained in way similar to
eq. (\ref{fund}) deriving explicitly the replicated free energy.
\begin{equation}
E=-\int P(J)dJ P_{no-J}(U,u_1,u_2)dU du_1 du_2 J \frac{\tanh(\beta
  U+\beta J)+\tanh\beta u_1 \tanh \beta u_2}{1+\tanh(\beta U+\beta
  J)\tanh\beta u_1 \tanh \beta u_2} 
\end{equation} 
From eqs. (\ref{Ppunto}) and (\ref{Pcoppia}) we see that the message
functions $Q(U,u_1,u_2)$ and $q(u)$ need not to be positive definite
but they must be such that $p(u)$,$P(U,u_1,u_2)$ and
$P_{no-J}(U,u_1,u_2)$ are. We will check that this is indeed the case
in the following.  In the last section we have argued that at the
critical temperature of the spin-glass phase transition the fields $u$
are small and at first order in $T-T_c$ the message functions are
described by the variables $(a,a_{11}(U),a_{12}(U))$ that represent
the average of the second moments of the fields, see eq. (\ref{eqA})
and (\ref{eqA1122}). Consistently we noted that these variables are
proportional to the eigenvector with zero eigenvalue of the Jacobian
matrix and checked that the eigenvector (and thus
$(a,a_{11}(U),a_{12}(U))$) is such that the function $Q(U,u_1,u_2)$
cannot be positive definite, see fig. \ref{figa11}.  In the following
we consider the similar quantities for $P(U,u_1,u_2)$ near the
critical temperature:
\begin{equation}
q_{11}(U)\equiv \int P(U,u_1,u_2)u_1^2du_1du_2
\end{equation}
\begin{equation}
q_{12}(U)\equiv \int P(U,u_1,u_2)u_1u_2du_1du_2
\end{equation}
Given that $P(U,u_1,u_2)$ \emph{is} a distribution we must find
$q_{11}(U)\geq 0$ and $q_{11}(U)-q_{12}(U)\geq 0$ for all $U$.  On the
other hand from eq. (\ref{Pcoppia}) we see that $q_{11}(U)$ and
$q_{12}(U)$ can be obtained from $(a,a_{11}(U),a_{12}(U))$ through a
linear transformation, again modulo an unknown positive constant that
scales as $T_c-T$. In fig. \ref{figq11} and \ref{figDQ2} we plot
$q_{11}(U)$ and $q_{11}(U)-q_{12}(U)$ for the 4DEA model with bimodal
interactions slightly below the critical temperature, they were
obtained through a linear transformation from the corresponding
$(a,a_{11}(U),a_{12}(U))$, ($a_{11}(U)$ is shown in
fig. \ref{figa11}).  Consistently we see that unlike $a_{11}(U)$, both
$q_{11}(U)$ and $q_{11}(U)-q_{12}(U)$ are positive for all $U$ as they
should.  One can check that similar quantities corresponding to
$P_{no-J}(U,u_1,u_2)$ or the plaquette are also compatible with the
general fact that the beliefs have to be interpreted as a distribution
functions over the disorder of the local effective Hamiltonians.

In the previous section we noticed that interpreting the message
function $Q(U,u_1,u_2)$ as a distribution of the messages
$(U,u_1,u_2)$ on a given instance is wrong and misleading. We have
seen why it is misleading: had we interpreted in that way we would
have look for a solution in the spin-glass phase with strictly
positive $a_{11}(U)$ while the true $a_{11}(U)$ is negative form some
values of $U$. We conclude this section providing some arguments to
show that it is inconsistent.

Consider the function for the belief of the single point
eq. (\ref{Ppunto}). Can we interpreter it as saying that $q(u)$
describes the distribution of the messages $u_i$ on a given
realization of the disorder? The answer is no because it would lead to
conclude that these messages are spatially uncorrelated. This is true
in the Bethe approximation but not in the plaquette approximation, for
instance the messages coming from the direction North and West come
from regions that are both contained in the NW plaquette and cannot be
uncorrelated. Furthermore we may notice that this equation remains the
same if we consider the CVM approximation in which the basic region is
a generic $L \times L$ plaquette. Once again on a given instance we
would have four messages entering on a point but when we average over
different samples these messages are correlated: thus it is crucial to
understand that the function $q(u)$ has nothing to do with the
distribution of the messages on a single sample. Actually the messages
are auxiliary objects of the approach while the true physical objects
are the beliefs.

Similarly looking at the equation for the beliefs of the couple of
points eq. (\ref{Pcoppia}) we see that if we interpreted
$Q(U,u_1,u_2)$ as the distribution over different samples of the
messages fields $(U,u_1,u_2)$ and $q(u)$ as that of $u$, we should
have concluded that the corresponding messages are spatially
uncorrelated and again we see that this is in contrast with the key
CVM assumption that objects in the same basic region (the plaquette)
are correlated.

Finally if we go back to the definition of the messages functions we
see that a generic message $\rho(\sigma_1,\sigma_2)$ in terms of
replicated spins must be positive function, instead the corresponding
function $Q(U,u_1,u_2)$ is some kind of integral transform of
$\rho(\sigma_1,\sigma_2)$ and need not to be positive.

We are now in position to discuss the relationship between the replica
CVM approach and the earlier results of the Tohoku group
\cite{T1,T2,T3}.  In 1980 Katsura, Fujiki and Nagahara were the first
to apply CVM ideas to spin-glasses and studied the phase diagrams of
various models.  They started from the CVM message-passing equations
on a given sample and introduced the functions $q(u)$ and
$Q(U,u_1,u_2)$ intended to be the sample-to-sample distributions of
the messages (actually they wrote $Q(U,u_1,u_2)=Q(U)g(u_1)g(u_2)$). As
we discussed in section \ref{plaquette}, these assumptions lead to the
same set of equations for $q(u)$ and $Q(U,u_1,u_2)$ that we have
obtained through Replica CVM at the RS level. However, even if the
equations are the same, the starting assumptions are inconsistent and
in the end the actual solutions $q(u)$ and $Q(U,u_1,u_2)$ turn out not
to be distributions.  Thus we think that in general Replica CVM
provides a more satisfactory and consistent derivation of the RS CVM
equations obtained nearly thirty years ago by the Tohoku group. Most
importantly the equations follow from a variational principle and can
be generalized to include RSB.

Actually Katsura {\it et al.}\ did not write down the full set of
equations (\ref{plav1},\ref{lhs},\ref{rhs}) but studied the
paramagnetic solution in various models and considered the location of
a second-order spin-glass phase transition. They solved equation
(\ref{PARAQU}) approximating $Q(U)$ as a couple of delta functions,
then in order to locate the transition temperature they studied the
Jacobian under the assumption that
$Q(U,u_1,u_2)=Q(U)g(u_1)g(u_2)$. This assumption is inconsistent with
the equation for $Q(U,u_1,u_2)$ that does not admit a factorized
solution but simplifies the computation of the Jacobian because one
has to deal with just two variables $a \equiv \int q(u)\,u^2\,du$ and
$a_{11} \equiv \int g(u)\,u^2\,du$ compared to the full set of
variables $(a,a_{11}(U),a_{12}(U))$ of the exact plaquette CVM. As
expected their results are less precise than ours, for instance they
obtained a spurious spin-glass phase transition on the 2DEA on the
square lattice while remarkably we find no second-order phase
transition.  On the other hand their treatment allows to obtain an
analytical expression of the Jacobian and of the phase diagrams and
they could apply it to various models.

\section{Discussion}

The Replica CVM can be used in principle to study the phase diagram of
a wide range of disordered systems and it is also possible that it can
be used to get quantitative predictions on single samples.  However
there are quantitative and qualitative technical differences with
respect to standard CVM (and the GBP algorithm), to the Bethe
approximation and to SP that represent a great challenge.  In
particular, integrals in many dimension, convolutions and message
functions that are not positive definite conjure to make applications
extremely difficult beyond the simplest CVM approximations and
smallest number of RSB.

Besides these technical difficulties the results of the second part of
the paper show that, at least in the averaged case, the method yields
sound results. This is not at all trivial given the frustrated nature
of the model and the tricky continuation in replica number.

In many respects the method shares the same advantages and weak points
of standard CVM.  It is a good tool to obtain quantitative
non-perturbative estimates in actual models and to characterize phase
diagrams, {\it e.g.}\ the stability of the paramagnetic phase down to
zero temperature in 2D spin-glass models is a remarkable result and we
already mentioned that the Jacobian approach can be straightforwardly
applied to study the ferromagnetic transition in $2D$ EA models with
ferromagnetically biased interactions.  Thus the method improves over
perturbative schemes, say the $1/D$ expansion that on the other hand
are more tractable in order to get qualitative results on systems with
full-RSB \cite{GMY}.

On the other hand a standard CVM approximation is not guaranteed to be
an upper bound to the true free energy and like many non-perturbative
approximations can possibly give inconsistent results, in this respect
we note that the plots of the free energy suggest that a good sequence
of Replica CVM approximations of increasing precision should approach
the true free energy from below and not from above as in standard
CVM. This appears to be a consequence of the replica trick much as the
fact that Parisi formula for the SK model has to be maximized and not
minimized in order to determine the free energy \cite{MPV}.  Another
drawback is that CVM is intrinsically a mean-field approximation. As
such it will always predict mean-field critical exponents although in
principle information on the true critical exponent can be obtained
\cite{PEL} {\it e.g.}\ comparing approximations with maximal regions
of increasing size.  However the problem of going beyond mean-field
theory ({\it i.e.}\ using the loop expansion in Replica Field Theory)
is way more important than to build more precise mean-field
approximations like those yielded by replica CVM.  This said
quantitative estimates can be rather useful to help solve
long-standing problems in spin-glass theory.  For instance an estimate
of the actual location of the De Almeida-Thouless line \cite{MPV} in
finite dimensional models is highly desirable.

The next step beyond the applications presented here should be the RS
treatment of the spin-glass phase of the EA model in the plaquette
approximation, at least in $D>3$. This include the extension to finite
small $n$ that can be seen as a first simplified version of 1RSB (the
so-called factorized solution). On the Bethe lattice \cite{GL} this
improves considerably the RS result and differs little from the more
precise 1RSB solution \cite{MP2}. A similar considerable improvement
of the factorized solution with respect to the RS one also in the
plaquette approximation, if observed, would constitute non-trivial
evidence supporting the RSB nature of the spin-glass phase in
finite-dimensional models below the upper critical dimension $6$.

A way to reduce the technical complexity of the exact Replica CVM
equations is to use the variational expression, eqs. (\ref{varf}) and
(\ref{e38}), of the free energy parametrizing the messages functions
in a simplified way and solving the corresponding variational
equations that will be different from the exact equations
(\ref{iteave1},\ref{iteave2}) and (\ref{itesamp1},\ref{itesamp2}).
The number of possible parametrizations is virtually infinite, for
instance it should be noted that in high-dimension the plaquette
approximation can be considered a perturbative correction to the Bethe
approximation and the RS function $Q(U,u_1,u_2)$ is peaked at small
values of all its arguments. Therefore in that regime it is consistent
to parametrize $Q(U,u_1,u_2)$ with few of its moments while
considering the full $q(u)$, the application of this ansatz to finite
dimension would be non-perturbative but could give consistent
results. A different possibility is to assume that in the whole
low-temperature phase, not only near $T_c$, the small fields
distributions are essentially parametrized by their second moments,
this amounts to consider as variables the functions $Q(U)$ and the
parameters $(a,a_{11}(U),a_{12}(U))$ introduced in section
\ref{sgtran}. To obtain the variational equations of this ansatz we
should parametrize the distribution of the fields $(U,u_1,u_2)$ as
delta functions plus second derivatives of delta functions, such that
all moments of order higher than two vanish.

The main reason to look for tractable replica CVM approximations
beyond the Bethe approximation is that one could then try to apply
them on single samples in the spirit of the SP algorithm. On the other
hand if one is just interested in the averaged properties it may be
useful to consider also tree-like approximation with regions of
increasing size. Being tree-like they are free of the convolutions and
negative message functions problems, {\it i.e.}\ fully treatable with
population dynamic algorithms. Therefore although they will be less
precise than CVM approximations larger regions could be treated and
extrapolation to the actual finite dimensional model achieved.

\begin{acknowledgements}
  We thank I. Nishimori, Y. Kabashima, J. Poulter and A. Pelizzola for
  useful comments on the first version of this work.  A. Lage and
  R. Mulet thank the hospitality of I.S.I. during the completion of
  this work.  A. Lage also acknowledges support by the EC-funded
  STREP GENNETEC (``Genetics Networks: emergence and copmlexity'')
\end{acknowledgements}

\appendix

\section{The Hierarchical Ansatz}
\label{RSBparametrization}

In this appendix we present the hierarchical ansatz for the replicated
CVM in full generality. In particular we will consider: i) a general
CVM approximation, ii) a general number of RSB steps, iii) both the
single sample and averaged cases. In appendix \ref{GSP} the
variational equations and the variational free energy will be derived.

In section \ref{CVMmp} we have written the free energy in terms of
messages that are positive functions of the configurations $x_r$ of
the variables in region $r$.  In the following we will parametrize the
messages in order to be able to take the analytical continuation to
real $n$, the RSB parametrization requires the introduction of a set
of $K$ RSB parameters $1 \leq x_1,\dots , x_K \leq n $ that are
numbers. Unfortunately at this stage the standard notations of the
message-passing formulation of CVM \cite{YFW} and of RSB \cite{MPV}
overlaps and the reader should not confuse the RSB parameter $1 \leq
x_1,\dots , x_K \leq n $ (that are numbers) with the $x_r$ defined
previously (that specifies distinct configurations of the variable
nodes in region $r$).  However from now on we will concentrate on the
Edwards-Anderson model defined above, therefore the nodes are Ising
spins, {\it e.g.}\ for the region comprising spin $a$ and $b$ we have
$x_{ab}\equiv \{\s_a,\s_b\}$ if the system is not replicated and
$x_{ab}\equiv \{\ts_a,\ts_b\}$ if the system is replicated $n$ times,
such that at each site, say $a$, we have $n$ spins $\ts_a \equiv
(\s_a^1, \dots \, \s_a^n)$.

The general RS and RSB ansatz of a function $\rho(\ts)$ of $n$ spins
was originally presented in \cite{GDD}, later its parametrization in
terms of distributions of fields was suggested in \cite{Mon1} and
later revisited in \cite{MP1}, and we refer to those paper for an
explanation of the main ideas underlying it. Here we generalize it to
a generic function $\rho(\ts_1,\dots ,\ts_p)$ where each $\ts_i$ is a
set of $n$ Ising spins.

We start introducing the field $U$ that parametrize a probability
distribution of $p$ Ising spins, in the following we call it a
$p$-field.  $U$ is a set of $2^p-1$ real numbers $\{u_I\}$ where $I$
is an index that labels all the subsets of the set of indices
$\{1,\dots,p\}$, (the empty set is excluded).  We have then:
\begin{equation}
P_{U}(\s_1,\dots,\s_p)= {\cal N}(U)  \exp \beta\left[\sum_i u_i \s_i +
  \sum_{i\leq j}u_{ij} \s_i \s_j+\sum_{i \leq j \leq k}u_{ijk}\s_i
  \s_j \s_k+\dots + u_{1,\dots,p}\s_1\dots\s_p \right] 
\label{uf}
\end{equation}   
Where ${\cal N}(U)$ is a normalization constant.  We define a
probability distribution (population) $P^{(0)}(U)$ of such fields a
$0$-distribution, correspondingly a $1$-distribution is a probability
distribution on probability distributions (population of populations)
and so on.  A $k$-distribution will be written as $P^{(k)}$ and it
defines a measure $P^{(k)}dP^{(k-1)}$ over the space of
$k-1$-distributions.

In order to parametrize the function $\rho(\ts_1,\dots ,\ts_p)$ with
$K$ steps of RSB we need:
\begin{itemize}
\item a $K$-distribution $P^{(K)}$ of the fields $U$;
\item $K$ integers $1 \leq x_1,\dots , x_K \leq n $ (for $n < 1$ they
  become real and the inequalities change sign);
\end{itemize}
In the following we will consider the parameters $1 \leq x_1,\dots ,
x_K \leq n $ fixed and consider just the dependency on the
distributions.  The construction is iterative and requires a set of
functions $\rho_{P^{(k)}}(\ts_1,\dots,\ts_p)$ where \emph{each $\ts_i$
  is a set of $x_{k+1}$ spins} with $k=1,\dots,K$ (we define $x_{K+1}
\equiv n$ and $x_0\equiv 1$). The normalization of $\rho_{P^{(k)}}$ is
crucial, we choose to normalize all of them to $1$.  We define
$\rho_{P^{(k)}}(\ts_1,\dots,\ts_p)$ starting from
$\rho_{P^{(k-1)}}(\ts_1,\dots,\ts_p)$ first dividing all the $x_{k+1}
\times p$ spins in $x_{k+1}/x_{k}$ groups $\{\ts_1,\dots,\ts_p\}_{\cal
  C}$ of $x_{k} \times p$ spins labeled by an index ${\cal C}=1,\dots
,x_{k+1}/x_{k}$. Then we have:
\begin{equation}
\rho_{P^{(k)}}(\ts_1,\dots,\ts_p )=\int P^{(k)}dP^{(k-1)}\prod_{{\cal
    C}=1}^{x_{k+1}/x_{k}}\rho_{P^{(k-1)}}(\{\ts_1,\dots,\ts_p\}_{\cal C}) 
\label{itdef}
\end{equation}
Thus $\rho(\ts_1,\dots,\ts_p)\equiv \rho_{P^{(K)}}(\ts_1,\dots,\ts_p)$
is defined iteratively starting from the Replica-Symmetric case
corresponding to $k=0$:
\begin{equation}
  \rho_{P^{(0)}}(\ts_1,\dots,\ts_p )=\int P^{(0)}(U)dU
  \prod_{i=1}^{x_1}P_{U}(\{\s_1,\dots,\s_p\}_i) 
\end{equation}
We will parametrize each message $m_{rs}(x_s)$ through a population
$P^{(K)}_{rs}$ over a $p$-field where $p$ is the number of sites in
region $s$. The populations associated to the tilded messages
$\tilde{m}_{rs}$ defined in eq. (\ref{tm}) will be represented by
$\tilde{P}^{(K)}_{rs}$.  Note that at integer values of $n$, the above
parametrization is redundant, and in principle we cannot determine the
population knowing the messages. It is standard in the replica method
to assume instead that this step can be performed due to the
continuation to real values of $n$.

In the end we will write down the message-passing equations in terms
of the message populations and thus we will be able to consider
non-integer values of $n$, and in particular the limit $n \rightarrow
0$.  We will also obtain a variational expression of the free energy
in terms of the message populations.  We note that in principle one
could parametrize the beliefs with populations and obtain an
expression of Kikuchi free energy in terms of these populations. The
resulting expression however is extremely complicated due to the
presence of the entropic terms in the form $\sum_{x_r}b_r \ln b_r$,
see {\it e.g.}\ \cite{Mon1} where the entropy of the point belief was
computed.  The main reason why we derived the variational expression
(\ref{varf}) in terms of the messages is precisely because it avoids
to deal with terms of the form $\sum_{x_r}b_r \ln b_r$.

The r.h.s. of the rescaled iteration equation (\ref{tm1}) defines a
function $\tilde{\bf m}_{rs}$ of the messages in $M(r)\setminus
M(s)$. Now since the messages are parametrized by $K$-populations, we
need to determine the corresponding function $\tilde{\bf
  P}_{rs}^{(K)}$ that yields a $K$-population as a function of the
$K$-populations of the messages, such that the iteration equation
translates into $\tilde{P}^{(K)}_{rs}=\tilde{\bf P}^{(K)}_{rs}$.

We start introducing two $J$-dependent functions $\tilde{\bf
  m}_{J,rs}$ and $\tilde{\bf N}_{J,rs}$ such that the following
equation is satisfied:
\begin{equation}
\tilde{\bf m}_{J,rs}\tilde{\bf N}_{J,rs}=\sum_{x_{r \setminus
    s}}\psi_{r\setminus s}^J (x_{r})\prod_{m_{r'' s''} \in
  M(r)\setminus M(s)}m_{r'' s''} 
\label{genJ}
\end{equation}
We use bold face for indicating functions of messages, but we stress
that while $\tilde{\bf m}_{J,rs}$ returns a normalized probability on
spins in region $s$, the function $\tilde{\bf N}_{J,rs}$ returns a
number which is precisely the normalization required by equation
(\ref{genJ}).  This equation can be written in terms of populations
and defines a $K$-population $\tilde{\bf P}^{(K)}_{J,rs}$ as a
function of the $K$-populations in $M(r)\setminus M(s)$.  It turns out
that the function $\tilde{\bf P}^{(K)}_{J,rs}$ can be defined in terms
of the function $\tilde{\bf P}^{(K-1)}_{J,rs}$ that in turn can be
defined through the function $\tilde{\bf P}^{(K-2)}_{J,rs}$ and so on.
The resulting expression for a given $k$ is:
\begin{equation}
\tilde{{\bf P}}^{(k)}_{J,rs}= \frac{1}{\tilde{{\bf  N}}^{(k)}_{J,rs}}
\int \left( \prod_{m_{r'' s''} \in M(r)\setminus M(s)} P_{r''
    s''}^{(k)} dP_{r'' s''}^{(k-1)} \right) [\tilde{ {\bf
    N}}^{(k-1)}_{J,rs}]^{x_{k+1}/x_k} \delta
(\tilde{P}^{(k-1)}_{rs}-\tilde{{\bf P}}^{(k-1)}_{J,rs}) 
\label{defP}
\end{equation}
where ${\tilde{{\bf N}}^{(k)}_{J,rs}}$ is a number that is a function
of the $k$-populations $P_{r'' s''}^{(k)}$ in $M(r)\setminus M(s)$
defined according to:
\begin{equation}
{\tilde{{\bf  N}}^{(k)}_{J,rs}} \equiv \int \left( \prod_{m_{r'' s''}
    \in M(r)\setminus M(s)} P_{r'' s''}^{(k)} dP_{r'' s''}^{(k-1)}
\right) [\tilde{ {\bf N}}^{(k-1)}_{J,rs}]^{x_{k+1}/x_k} \ . 
\label{defN}
\end{equation}
In the cavity formulation at the Bethe level \cite{MP1} the quantities
${\bf N}^{(k)}_{J,rs}$ are associated to (cluster) free energy shifts
and they appear in (\ref{defP}) as reweighting terms.  At the Bethe
level, explicit expressions like (\ref{defP}) for any $k$ value have
been already reported in Ref.~\cite{GSfreezing}.

The iterative definition must be supplemented with the two functions
for $k=0$ and $k=-1$.  They reads:
\begin{equation}
\tilde{{\bf P}}^{(0)}_{J,rs}= \frac{1}{\tilde{{\bf  N}}^{(0)}_{J,rs}}
\int \left( \prod_{m_{r'' s''} \in M(r)\setminus M(s)} P_{r''
    s''}^{(0)} dU_{r'' s''} \right) [\tilde{ \bf
  N}^{(-1)}_{J,rs}]^{x_1} \delta (\tilde{U}_{rs}-\tilde{{\bf
    U}}_{J,rs}) 
\label{defP0}
\end{equation}
and 
\begin{equation}
{\tilde{{\bf  N}}^{(0)}_{J,rs}} \equiv \int \left( \prod_{m_{r'' s''}
    \in M(r)\setminus M(s)} P_{r'' s''}^{(0)} dU_{r'' s''} \right)
[\tilde{ {\bf N}}^{(-1)}_{J,rs}]^{x_{1}}  
\label{defN0}
\end{equation}
For $x_1=n$ this yields the Replica-Symmetric solution, note that in
this case in the $n \rightarrow 0$ limit the reweighting term goes to
$1$ and is irrelevant.  In the above equations $\tilde{{\bf U}}_{rs}$
and $\tilde{ {\bf N}}^{(-1)}_{rs}$ are functions of the fields $U_{r''
  s''}$ that have to be obtained solving the following single-replica
equation:
\begin{equation}
\rho_{\tilde{\bf U}_{J,rs}} ({\s}_s) {\bf N}_{J,rs}^{(-1)} =
\sum_{\{\s_{r \setminus  s}\}}\psi_{r\setminus s}^J
(\s_r)\prod_{m_{r'' s''} \in M(r)\setminus M(s)}\rho_{{U}_{r''
    s''}}(\s_{s''}) 
\label{obb2}
\end{equation}   
Note that it is only at this stage that the actual properties of the
model enter.  This complete the recursive definition of the function
$\tilde{{\bf P}}^{(k)}_{J,rs}$ and $\tilde{\bf N}_{J,rs}^{(k)}$that
solve equation (\ref{genJ}) at any level of $k$-RSB, we will not write
down the proof of this statement that can be worked out iteratively
generalizing a similar derivation in the Bethe approximation
\cite{PRdil}.  It is important to stress that this iterative
definition is possible only if we do not average over the disorder, a
passage that will be taken in the next section.
  
Up to now we have expressed eq. (\ref{tm1}) in terms of populations in
such a way that the $n \rightarrow 0$ limit can be taken, in the
following we consider the similar treatment for eq. (\ref{tm}).  Note
the basic differences between eq. (\ref{tm}) and (\ref{tm1}): the
absence of the summation over the spins in region $r\setminus s$ and
the fact that we consider messages in $M(r,s)$ and not in
$M(r)\setminus M(s)$.

Much as above, the function $\tilde{{\bf Q}}^{(k)}_{rs}$ corresponding
to the r.h.s. of eq. (\ref{tm}) can be defined in an recursive way,
the result being:
\begin{equation}
\tilde{{\bf Q}}^{(k)}_{rs}= \frac{1}{\tilde{{\bf  M}}^{(k)}_{rs}} \int
P_{rs}^{(k)} dP_{rs}^{(k-1)}\left( \prod_{m_{r'' s''} \in M(r,s)}
  P_{r'' s''}^{(k)} dP_{r'' s''}^{(k-1)} \right) [\tilde{ {\bf
    M}}^{(k-1)}_{rs}]^{x_{k+1}/x_k} \delta
(\tilde{P}^{(k-1)}_{rs}-\tilde{{\bf Q}}^{(k-1)}_{rs}) 
\label{defQ}
\end{equation}
where ${\tilde{{\bf M}}^{(k)}_{rs}}$ is defined as:
\begin{equation}
{\tilde{{\bf  M}}^{(k)}_{rs}} \equiv \int P_{rs}^{(k)}
dP_{rs}^{(k-1)}\left( \prod_{m_{r'' s''} \in M(r,s)} P_{r'' s''}^{(k)}
  dP_{r'' s''}^{(k-1)} \right) [\tilde{ {\bf
    M}}^{(k-1)}_{rs}]^{x_{k+1}/x_k} \ . 
\label{defM}
\end{equation}
As before the previous iterative definition is completed specifying
the two function at the Replica-Symmetric level corresponding to
$k=0$:
\begin{equation}
\tilde{{\bf Q}}^{(0)}_{rs}= \frac{1}{\tilde{{\bf { M}}}^{(0)}_{rs}}
\int P_{rs}^{(0)} dU_{rs}\left( \prod_{m_{r'' s''} \in M(r,s)} P_{r''
    s''}^{(0)} dU_{r'' s''} \right) [\tilde{ {\bf
    M}}^{(-1)}_{rs}]^{x_1} \delta (\tilde{U}_{rs}-\tilde{{\bf
    Q}}_{rs}^{(-1)}) 
\label{defQ0}
\end{equation}
and 
\begin{equation}
{\tilde{{\bf { M}}}^{(0)}_{rs}} \equiv \int P_{rs}^{(0)} dU_{rs}\left(
  \prod_{m_{r'' s''} \in M(r,s)} P_{r'' s''}^{(0)} dU_{r'' s''}
\right) [\tilde{ {\bf M}}^{(-1)}_{rs}]^{x_{1}} \ . 
\label{defM0}
\end{equation}
For $x_1=n$ this gives back the Replica-Symmetric solution, note that
in this case in the $n \rightarrow 0$ limit the reweighting term goes
to $1$.

The two quantities $\tilde{{\bf Q}}_{rs}^{(-1)}$ and $\tilde{ {\bf
    M}}^{(-1)}_{rs}$ have the following form that does not depend on
the Hamiltonian of the problem:
\begin{equation}
{\bf Q}_{rs}^{(-1)} \equiv U_{rs}+\sum_{m_{r''s''} \in M(r,s)} U_{r''s''}
\label{uuu}
\end{equation} 
The above sum is intended in vectorial form, {\it i.e.}\ for each
possible combination of the spins in region $s$
($\s_i$,\,$\s_i\s_j$,\,$\s_i\s_j\s_k$\,\dots) we sum the corresponding
fields.  The normalization factor can be written as the product of
terms depending on each message separately times a term depending on
the sum of the fields:
\begin{equation}
{\bf M}_{rs}^{(-1)}  \equiv \frac{{\cal N}(U_{rs})\prod_{m_{r''s''}
    \in M(r,s)} {\cal N}(U_{r''s''})  }{{\cal N}({\bf Q}_{rs}^{(-1)})} 
\label{mmm}
\end{equation} 
The above properties have important consequences and allow to
introduce Fourier-like transforms in order to write down the equations
for the messages in explicit form, see appendix \ref{inversion}.

\section{Generalized Survey Propagation Equations}
\label{GSP}

We are now in position to write the variational equations
(\ref{yedmessage}) in terms of $K$-populations.  We start noticing
that in the averaged case eq. (\ref{tm1}) can be written as:
\begin{equation}
\tilde{m}_{rs} C = \langle \tilde{\bf m}_{J,rs} \tilde{\bf N}_{J,rs}\rangle_J
\label{mC}
\end{equation}
where $C$ is a normalization constant.  The r.h.s. of the previous
equation is a linear combination of spin probabilities and it is easy
to see that the corresponding $K$-population is just a linear
combination of the corresponding $K$-populations with same
coefficients $\tilde{\bf N}_{J,rs}$ and the normalization factor is
just that sum of the coefficients i.e. $\langle \tilde{\bf N}_{J,rs}
\rangle_J$.  The populations corresponding to $\tilde{\bf m}_{J,rs}$
are given by eq. (\ref{defP}) and we see that the coefficients
$\tilde{\bf N}_{J,rs}$ in eq. (\ref{mC}) cancel the term at the
denominator in (\ref{defP}), thus the resulting equation for the
populations $P^{(K)}_{rs}$ is:
\begin{equation}
\tilde{P}^{(K)}_{rs}=\frac{1}{\langle \tilde{{\bf
      N}}^{(K)}_{J,rs}\rangle_J} \int \left( \prod_{m_{r'' s''} \in
    M(r)\setminus M(s)} P_{r'' s''}^{(K)} dP_{r'' s''}^{(K-1)}
\right)\langle [\tilde{ {\bf N}}^{(K-1)}_{J,rs}]^{x_{K+1}/x_K} \delta
(\tilde{P}^{(K-1)}_{rs}-\tilde{{\bf P}}^{(K-1)}_{J,rs}) \rangle_J 
\label{iteave1}
\end{equation}  
\begin{equation}
\tilde{P}^{(K)}_{rs}= \tilde{\bf Q}^{(K)}_{rs} 
\label{iteave2}
\end{equation}
Note that in the limit $n \rightarrow 0$ we have $\tilde{{\bf
    N}}^{(K)}_{rs} \rightarrow 1$ because $x_{K+1}=n$.  These
equations are translationally invariant, meaning that the populations
$P^{(K)}_{rs}$ does not depends on where regions $r$ and $s$ are
actually on the lattice but just on their shape and mutual positions
with respect to each other.  For instance in the Bethe approximation
we have a single $K$-population that does not fluctuate over the
sites.  The right hand sides of the above equations (\ref{iteave1})
and (\ref{iteave2}) should be thought of as respectively the
r.h.s. and l.h.s. of a single equation that in general involves the
messages populations $P^{(K)}_{rs}$ in implicit form.  This is the
main difference with respect to the Bethe approximation where the
messages appears in explicit form and the equations can be solved
iteratively, a discussion of these equations in the context of the CVM
plaquette approximation will be given at the end of next section.

On a single sample instead the distributions fluctuate over the
sites. As a consequence the solution with $K$-RSB steps is
parametrized by $(K-1)$-RSB populations fluctuating over space. The
corresponding equations are:
\begin{equation}
\tilde{P}^{(K-1)}_{rs}= \tilde{\bf P}^{(K-1)}_{J,rs}
\label{itesamp1}
\end{equation} 
\begin{equation}
\tilde{P}^{(K-1)}_{rs}= \tilde{\bf Q}^{(K-1)}_{rs}
\label{itesamp2}
\end{equation} 
For instance the $1$-RSB solution on a given sample is described by
populations of fields that fluctuate over different regions.  Again
the right hand sides of the above equations (\ref{itesamp1}) and
(\ref{itesamp2}) should be thought of as respectively the r.h.s. and
l.h.s. of a single equation that in general involves the messages in
implicit form.

When $K=1$ and the maximal region is the couple of points ({\it
  i.e.}\ we work in the Bethe approximation) the above equations
reproduce the celebrated Survey Propagation (SP) equations
\cite{MP1,MP2,MZ,MZP}, hence the name Generalized Survey Propagation
equations.  The SP equations were originally obtained through the
cavity method, while a derivation of the SP equations on a single
sample using replicas was obtained in \cite{KABA}.

\subsection{The Free Energy and its Derivatives}

The variational expression of the free energy eq. (\ref{varf}) can be
written in terms of populations along the same lines of the previous
sections.  The results in the averaged case is:
\begin{equation}
F_K=-\sum_{r \in R}c_r \ln \langle {\bf N}^{(K)}_{J,r} \rangle_J
\end{equation}
where ${\bf N}^{(K)}_{J,r}$ is a $J$-dependent function of the
$K$-Populations corresponding to the messages in $M(r)$ that can be
defined iteratively in terms of the populations as:
\begin{equation}
{\bf N}^{(k)}_{J,r}=\int \prod_{r''s'' \in M(r)}
P_{r''s''}^{(k)}dP_{r''s''}^{(k-1)}({\bf
  N}^{(k-1)}_{J,r})^{x_{k+1}/x_k} 
\end{equation}
with 
\begin{equation}
{\bf N}^{(0)}_{J,r}=\int \prod_{r''s'' \in M(r)}
P_{r''s''}^{(0)}dU_{r''s''}({\bf N}^{(-1)}_{J,r})^{x_{1}} 
\end{equation}
where 
\begin{equation}
{\bf N}^{(-1)}_{J,r}=\sum_{\{\s_{r}\}}\psi_{r}^J (\s_r)\prod_{r'' s''
  \in M(r)}\rho_{{U}_{r'' s''}}(\s_{s''}) 
\end{equation}
On a single instance instead we have:
\begin{equation}
F_K=-\sum_{r \in R}c_r \ln  {\bf N}^{(K-1)}_{J,r} 
\end{equation}
where the $ {\bf N}^{(K-1)}_{J,r}$ have to be evaluated in terms of
the $K-1$-populations entering region $r$.  In order to make explicit
the similarity with the cavity formulation of \cite{MP1} we define a
new function
\begin{equation}
{\bf \Delta F}^{(k)}_{J,r} \equiv-\frac{1}{\beta x_{k+1}}\ln{\bf N}^{(k)}_{J,r}
\end{equation} 
Consequently the above quantity has again a recursive definition:
\begin{equation}
{\bf \Delta F}^{(k)}_{J,r}=-{1 \over \beta x_{k+1}}\ln \int
\prod_{r''s'' \in M(r)} P_{r''s''}^{(k)}dP_{r''s''}^{(k-1)}e^{-\beta
  x_{k+1}{\bf \Delta F}^{(k-1)}_{J,r}} 
\end{equation}
with this definition we have:
\begin{equation}
\Phi(n)=-{1 \over \beta n \, N}\sum_{r \in R}c_r \ln \langle e^{-\beta
  n{\bf \Delta F}^{(K)}_{J,r}} \rangle_J 
\label{e38}
\end{equation}
and 
\begin{equation}
f=\lim_{n \rightarrow 0}\Phi(n)={1 \over N}\sum_{r \in R}c_r \int
\prod_{r''s'' \in M(r)} P_{r''s''}^{(K)}dP_{r''s''}^{(K-1)}\langle{\bf
  \Delta F}^{(K-1)}_{J,r}\rangle_J 
\label{e39}
\end{equation}
In the average case the message populations do not fluctuate over
space and the contribution of each region of a given form to the free
energy is the same. Thus the sum over all regions can be replaced by
the sum of the contributions of the basic types of regions $r$ each
multiplied by the number $n_r$ of regions of type $r$ per spin.  For
instance in the plaquette approximation of the 2D square lattice we
have $c_\text{plaquette}=1$, $c_\text{couple}=-1$ and
$c_\text{point}=1$, and the number of regions per spin are
respectively $n_\text{plaquette}=1$, $n_\text{couple}=2$ and
$n_\text{point}=1$.

It is also important to determine the derivatives of the free energy
with respect to the parameters $1 \leq x_1,\dots , x_K \leq n $
\cite{MP1,MP2}. Since the above expression is variational with respect
to the populations the total derivative of the free energy with
respect to $x_k$ coincides with its partial derivative with respect to
$x_k$.

Clearly in order to determine the derivative of the free energy with
respect to the parameters $1 \leq x_1,\dots , x_K \leq n $ we only
need to determine the derivatives of the function ${\bf \Delta
  F}^{(K)}_{J,r}$.  The derivative of ${\bf \Delta F}^{(p)}_{J,r}$
with respect to $x_{p+1}$ is:
\begin{equation}
\partial_{x_{p+1}}{\bf \Delta F}^{(p)}_{J,r}=-{1 \over x_{p+1}}{\bf
  \Delta F}^{(p)}_{J,r}+\langle\langle {\bf \Delta
  F}^{(p-1)}_{J,r}\rangle\rangle^{(p)}_{J,r} 
\end{equation}
where we have defined: 
\begin{equation}
\langle \langle \dots  \rangle\rangle^{(k)}_{J,r}=\frac{\int
  \prod_{r''s'' \in M(r)} P_{r''s''}^{(k)}dP_{r''s''}^{(k-1)}\dots
  e^{-\beta x_{k+1}{\bf \Delta F}^{(k-1)}_{J,r}}}{\int \prod_{r''s''
    \in M(r)} P_{r''s''}^{(k)}dP_{r''s''}^{(k-1)}e^{-\beta x_{k+1}{\bf
      \Delta F}^{(k-1)}_{J,r}}} 
\end{equation}
Then the derivative $\partial_{x_{p+1}}{\bf \Delta F}^{(K)}_{J,r}$ can
be obtained using the recursive definition.  Indeed we see that the
derivatives of ${\bf \Delta F}^{(k)}_{J,r}$ for $k>p$ with respect to
$x_{p+1}$ parameters obey the following recursive equation:
\begin{equation}
\partial_{x_{p+1}}  {\bf \Delta F}^{(k)}_{J,r}=\langle
\langle \partial_{x_{p+1}}  {\bf \Delta F}^{(k-1)}_{J,r}
\rangle\rangle^{(k)}_{J,r} \ \ \ \ {\rm for} \ \ k>p 
\end{equation}
For instance in the 1RSB case in the $n \rightarrow 0$ limit we get a
similar result to Ref.~\cite{MP1}:
\begin{equation}
{\partial f_{1RSB} \over \partial x_1}=\sum_{r \in R}c_r \int
\prod_{r''s'' \in M(r)} P_{r''s''}^{(1)}dP_{r''s''}^{(0)}\left(-{1
    \over x_{1}}\langle{\bf \Delta F}^{(0)}_{J,r}\rangle_J+ \langle \
  \langle\langle {\bf \Delta
    F}^{(-1)}_{J,r}\rangle\rangle^{(0)}_{J,r}\ \rangle_J\right) 
\end{equation}

\subsection{From populations to functionals}

In the preceding sections we have worked under the assumption that the
messages and the beliefs are both parametrized through populations of
populations of fields.  In section \ref{plaquette} we have applied the
approach to the EA model on various lattices in the CVM plaquette
approximation.  We made the simplest non-trivial ansatz, {\it
  i.e.}\ the RS ansatz, that we expect to be good in the
high-temperature paramagnetic phase.  On a single sample the messages
are just numbers in the RS phase, and the corresponding equations are
the GBP equations of \cite{YFW}. For the replicated averaged system
instead the messages according to the above parametrization should be
distributions of fields. As we have seen it turns out that the actual
solution requires to consider messages parametrized by functions that
are not positive definite {\it i.e.}\ that are not distributions.
This tells us that in practice when solving a given model we may be
forced to consider parametrization of the form (\ref{itdef}) where
instead of populations of fields we should consider functions of the
fields, or instead of populations of populations we should consider
functionals. At any rate, it is easily seen that the equations obeyed
by these objects do not change at all depending on whether they are
distributions or functions {\it i.e.}\ all the results of the previous
subsection hold. On the other hand it can be argued that the function
$\tilde{N}^{(k)}_{J,rs}$ defined above should always give a positive
number and this is guaranteed only if it is a function of
$k$-populations, therefore it appears that in any case we should
parametrize a message at level $K$ of RSB with a function of
$K-1$-populations and this function may eventually be non-positive
definite. In other words it seems to us that we can relax the
condition to use distributions only at the last level of RSB.

\section{The Inversion Problem}
\label{inversion}

The generalized survey propagation equations derived previously
express the tilded populations in terms of the non-tilded ones.  To
obtain an explicit expression for the message functions one needs to
invert equations (\ref{iteave2}) (or eq. (\ref{itesamp2}) on a given
sample) and obtain an expression for the non-tilded populations in
terms of the tilded ones.  At the lowest RSB levels, ({\it i.e.}\ RS
in the averaged case and 1RSB on a single sample) this can be done
using Fourier transforms.  In this appendix we will show that at any
number $K$ of RSB steps the inversion can be achieved in principle
using appropriate integral transforms of the populations.  An
algorithm able to go back and forth from the populations to their
transforms would provide a route to the numerical solution of the
Generalized Survey Propagation equations but unfortunately an
efficient implementation seems quite difficult.

Basically we will work with nested Fourier transform, {\it
  i.e.}\ Fourier transforms of Fourier transforms.  We will introduce
the invertible integral transform ${\cal T}^{(k)}_{rs}$ of a
$k$-population $P^{(k)}_{rs}$ and show that eq. (\ref{iteave2}) in
terms of transform reads:
\begin{equation}
\tilde{\cal T}_{rs}^{(k)}={\cal T}^{(k)}_{rs}+\sum_{m_{r''s''} \in
  M(r,s)}{\cal T}^{(k)}_{r''s''} 
\label{eqt1}
\end{equation}
From the previous expression we can easily compute ${\cal
  T}^{(k)}_{rs}$ from $\tilde{\cal T}_{rs}^{(k)}$.  Again we will
proceed iteratively, showing that if the previous equation is valid at
level $k-1$ it is also valid at level $k$.  In order to do this we
also assume that it exists a function ${\cal G}^{(k-1)}$ such that the
quantity $\tilde{ {\cal M}}^{(k-1)}_{rs}$ can be written as
\begin{equation}
\tilde{ {\cal M}}^{(k-1)}_{rs} = \frac{{\cal G}^{(k-1)}({\cal
    T}^{(k-1)}_{rs})\prod_{m_{r''s''} \in M(r,s)} {\cal
    G}^{(k-1)}({\cal T}^{(k-1)}_{r''s''})}{{\cal
    G}^{(k-1)}(\tilde{\cal T}^{(k-1)}_{rs})} 
\label{eqt2}
\end{equation}  
where $\tilde{\cal T}^{(k-1)}_{rs})$ in the above equation is computed
as the sum of the non-tilded transform.  We can write eq. (\ref{defQ})
in terms of the transforms at level $k-1$:
\begin{eqnarray}
\tilde{{\bf Q}}^{(k)}_{rs}(\tilde{\cal T}^{(k)}_{rs}) (\tilde{\cal
  T}^{(k-1)}_{rs}) & = & \frac{1}{\tilde{{\bf {\cal M}}}^{(k)}_{rs}}
\int P_{rs}^{(k)}({\cal T}_{rs}^{(k-1)}) d{\cal T}_{rs}^{(k-1)}\left(
  \prod_{m_{r'' s''} \in M(r,s)} P_{r''s''}^{(k)}({\cal
    T}_{r''s''}^{(k-1)}) d{\cal T}_{r''s''}^{(k-1)}\right) [\tilde{
  {\cal M}}^{(k-1)}_{rs}]^{x_{k+1}/x_k}  
\times 
\non
\\
& \times &
\delta \left(\tilde{\cal T}^{(k-1)}_{rs}-{\cal
    T}^{(k-1)}_{rs}-\sum_{m_{r''s''} \in M(r,s)}{\cal
    T}^{(k-1)}_{r''s''} \right) 
\end{eqnarray}
now using eq. (\ref{eqt2}) we can redistribute the reweighting term
between the various populations and rewrite the previous equation as:
\begin{eqnarray}
\tilde{{\bf Q}}^{(k)}_{rs} (\tilde{\cal G}_{rs}^{(k-1)})^{x_{k+1}/x_k}
& = & \frac{1}{\tilde{{\bf {\cal M}}}^{(k)}_{rs}} \int P_{rs}^{(k)}
({\cal G}_{rs}^{(k-1)})^{x_{k+1}/x_k}d{\cal T}_{rs}^{(k-1)}\left(
  \prod_{m_{r'' s''} \in M(r,s)} P_{r''s''}^{(k)}({\cal
    G}_{r''s''}^{(k-1)})^{x_{k+1}/x_k} d{\cal
    T}_{r''s''}^{(k-1)}\right)  
\times
\non
\\
& \times &
\delta \left(\tilde{\cal T}^{(k-1)}_{rs}-{\cal
    T}^{(k-1)}_{rs}-\sum_{m_{r''s''} \in M(r,s)}{\cal
    T}^{(k-1)}_{r''s''} \right) 
\end{eqnarray}
In the previous equation $\tilde{{\bf Q}}^{(k)}_{rs} (\tilde{\cal
  G}_{rs}^{(k-1)})^{x_{k+1}/x_k}$ defines a measure over the space of
$k-1$ transform. We Fourier transform the previous equation with
respect to the $k-1$ transform and obtain:
\begin{equation}
\tilde{\cal F}^{(k)}_{rs}=\frac{1}{\tilde{{\bf {\cal M}}}^{(k)}_{rs}}
{\cal F}_{rs}^{(k)} \prod_{m_{r'' s''} \in M(r,s)} {\cal F}_{r''s''}^{(k)}
\end{equation}
accordingly the factor $\tilde{{\bf {\cal M}}}^{(k)}_{rs}$ can be
obtained as the transform evaluated at a given value, {\it e.g.}\ zero
argument; this yields:
\begin{equation}
\tilde{{\bf {\cal M}}}^{(k)}_{rs} =\frac{\tilde{\cal F}_{rs}^{(k)}(0)
\prod_{m_{r'' s''} \in M(r,s)} {\cal F}_{r''s''}^{(k)}(0)}{\tilde{\cal
  F}^{(k)}_{rs}(0)} 
\end{equation}
Now we can take the logarithm of the previous equation and define:
\begin{equation}
\tilde{\cal T}^{(k)}_{rs} \equiv \ln \frac{\tilde{\cal
    F}_{rs}^{(k)}}{\tilde{\cal F}_{rs}^{(k)}(0)}  
\label{defT}
\end{equation}
we see that with this definition eq. (\ref{eqt1}) is satisfied at
level $k$.  Correspondingly eq. (\ref{eqt2}) is also satisfied
provided we identify the function $\cal G$ with:
\begin{equation}
\tilde{\cal G}_{rs}^{(k)} \equiv \tilde{\cal F}_{rs}^{(k)}(0)
\end{equation}
To complete the proof we need to show that eq. (\ref{eqt1}) and
eq. (\ref{eqt2}) are valid at $k=0$, the replica symmetric case.  To
do this we easily see that eqs. (\ref{uuu}) and (\ref{mmm}) have this
structure. Therefore at the RS level the transform ${\cal
  T}^{(0)}_{rs}$ is defined as:
\begin{equation}
{\cal T}_{(rs)}^{(0)}(S_{rs})=\ln \frac{\int P^{(0)}_{rs} (U_{rs})
  {\cal N}(U_{rs})^{x_1}d U_{rs} \exp i S_{rs} U_{rs}}{\int
  P^{(0)}_{rs}(U_{rs})  {\cal N}(U_{rs})^{x_1} dU_{rs} } 
\end{equation}
and we have:
\begin{equation}
{\cal G}^{(0)}_{rs}=\int P^{(0)}_{rs}(U_{rs})  {\cal N}(U_{rs})^{x_1} dU_{rs} 
\end{equation}
Summarizing, at the RS level, the integral transform needed to solve
the inversion problem is defined as the logarithm of the Fourier
transform of the reweighted distribution of fields.  Note that if the
solution is purely RS we have $x_1=n$ and the reweighting term is
absent in the $n \rightarrow 0$ limit.  From the above RS expression
one can go to 1RSB using eq. (\ref{defT}). At 1RSB the solution is
parametrized by a population of populations, and the transform is the
logarithm of the Fourier transform of the reweighted distribution of
the logarithm of the Fourier transforms of the populations.

\end{document}